\documentclass[conference,compsoc]{IEEEtran}

\usepackage[nocompress]{cite}

\usepackage{url}

\makeatletter
\g@addto@macro{\UrlBreaks}{\UrlOrBreaks}
\makeatother

\def\UrlBreaks{\do\a\do\b\do\c\do\d\do\e\do\f\do\g\do\h\do\i\do\j%
\do\k\do\l\do\m\do\n\do\o\do\p\do\q\do\r\do\s\do\t\do\u\do\v\do\w\do\x\do\y\do\z%
\do\A\do\B\do\C\do\D\do\E\do\F\do\G\do\H\do\I\do\J\do\K\do\L\do\M\do\N\do\O%
\do\P\do\Q\do\R\do\S\do\T\do\U\do\V\do\W\do\X\do\Y\do\Z%
\do\0\do\1\do\2\do\3\do\4\do\5\do\6\do\7\do\8\do\9%
\do\-\do\.\do\?\do\=\do\/\do\&}

\usepackage[breaklinks=true,colorlinks=true,allcolors=blue,pdfborder={0 0 0}]{hyperref}

\usepackage[utf8]{inputenc}

\usepackage{tikz}
\usepackage{amsmath}

\usepackage{filecontents}

\usepackage{caption,tabularx,booktabs}
\usepackage{pifont}
\usepackage{algorithm}
\usepackage{algpseudocode}

\usepackage{xcolor,colortbl}
\usepackage{tikz}
\usepackage{multirow}

\usepackage{wrapfig}

\newcommand{\RV}{\textit{RVDebloater}}

\newcommand{\DT}{adaptive}
\newcommand{\RVC}{\textit{RVDebloater-static}}
\newcommand{\RVA}{\textit{RVDebloater-dynamic}}

\usepackage{listings}
\usepackage{makecell}
\usepackage{arydshln}
\usepackage{enumerate}
\usepackage[shortlabels]{enumitem}
\usepackage{diagbox}
\usepackage{rotating}
\usepackage{subcaption}
\usepackage{float}

\newcommand{\Input}{\State\textbf{inputs:}\ }
\newcommand{\Output}{\State\textbf{output:}\ }

\definecolor{commentsColor}{rgb}{0, 0.5, 0}
\definecolor{keywordsColor}{rgb}{0.000000, 0.000000, 0.635294}
\definecolor{stringColor}{rgb}{0.558215, 0.000000, 0.135316}
\definecolor{applegreen}{rgb}{0.55, 0.71, 0.0}

\lstdefinelanguage{myLang}
{
	morekeywords={
		bool,
    u32_int
	},
	sensitive=false, 
	morecomment=[l]{//}, 
	morecomment=[s]{/*}{*/}, 
	morestring=[b]" 
}
\lstdefinelanguage{myLangLL}
{
	morekeywords={
		entry,
		alloca,
		\define,
		i32,
		i64,
		call,
		void
	},
	sensitive=false, 
	morecomment=[l]{//}, 
	morecomment=[s]{/*}{*/}, 
	morestring=[b]" 
}

\definecolor{greenannoback}{RGB}{230,244,214}

\begin{document}
    
\title{RVDebloater: Mode-based Adaptive Firmware Debloating for Robotic Vehicles}

\author{\IEEEauthorblockN{Mohsen Salehi}
\IEEEauthorblockA{The University of British Columbia \\
Vancouver, Canada\\
msalehi@ece.ubc.ca}
\and
\IEEEauthorblockN{Karthik Pattabiraman}
\IEEEauthorblockA{The University of British Columbia\\
Vancouver, Canada\\
karthikp@ece.ubc.ca}}

\maketitle

\begin{abstract}
As the number of embedded devices grows and their functional requirements increase,
  embedded firmware is becoming increasingly larger, thereby expanding its attack surface.
  Despite the increase in firmware size, many embedded devices, such as robotic vehicles (RVs), 
  operate in distinct modes, 
  each requiring only a small subset of the firmware code at runtime.
  We refer to such devices as \textit{mode-based embedded devices}.
Debloating is an approach to reduce attack surfaces by removing or restricting unneeded code, 
but existing techniques suffer from significant limitations, such as coarse granularity and irreversible code removal, 
limiting their applicability.

  To address these limitations, we propose \RV{}, a novel adaptive debloating technique for 
  \textit{mode-based embedded devices} 
  that automatically identifies unneeded firmware code for each mode 
  using either static or dynamic analysis,
  and dynamically debloats 
  the firmware for each mode at the function level at runtime. 
  \RV{} introduces a new software-based enforcement approach that supports diverse 
  \textit{mode-based embedded devices}. 
  We implemented \RV{} using the LLVM compiler and evaluated its efficiency and effectiveness
  on six different RVs, including both simulated and real ones, with different real-world missions.
  We find that device requirements change throughout its lifetime for each mode, 
  and that many critical firmware functions can be restricted in other modes, 
  with an average of 85\% 
  of functions not being required.
  The results showed that none of the missions failed after debloating with \RV{}, indicating that it neither incurred false positives nor false negatives. 
  Further, \RV{} prunes the firmware call graph by an average of 45\% across different firmware. 
  Finally, \RV{} incurred an average performance overhead of 3.9\% and memory overhead of 4\% (approximately 0.25 MB) on real RVs. 
\end{abstract}

\section{Introduction}
\label{sec-intro}
Robotic vehicles (RVs) such as unmanned aerial vehicles (UAVs), 
unmanned ground vehicles (UGVs), and remotely operated underwater vehicles (ROVs), are a prominent class of embedded devices.
A wide range of features (e.g., system calls and device drivers~\cite{huang2022ksplit}) 
and hardware functionalities is being incorporated into these devices 
to support the requirements of different users, 
even though each user may only require a fraction of these features.
This expansion 
enhances potentially buggy code, 
even in unused features, thereby increasing the devices' attack surface~\cite{hooper2016securing,HijackD,quarta2017experimental}.
For instance, 
UAVs are vulnerable to various security vulnerabilities, 
such as buffer overflows~\cite{kim2018securing, hooper2016securing}, which can allow attackers to hijack the control flow 
and manipulate critical programs like the flight control program, causing crashes. 


One approach to reduce attack surfaces is minimizing 
or restricting unwanted features (e.g., security-critical system calls) through a technique known as \emph{debloating}~\cite{azad2019less,quach2018debloating,kim2018securing,hu2023hacksaw,hu2022irqdebloat,gu2014face,heo2018effective,abubakar2021shard}. 
Debloating \emph{keeps} a \emph{small subset} of firmware code necessary for specific workloads 
while \emph{restricting} the execution of the remaining code 
\cite{ghavamnia2020temporal,abubakar2021shard,rajagopalan2023syspart,kim2018securing},
or \emph{removes} unwanted features such as peripherals (e.g., USB and UART~\cite{hu2022irqdebloat,hu2023hacksaw}). 
Although these techniques have gained popularity in different contexts, 
such as desktop~\cite{abubakar2021shard,gu2014face} and 
server applications~\cite{rajagopalan2023syspart,jahanshahi2023minimalist,azad2019less}, 
embedded systems have not received sufficient attention~\cite{kim2018securing}.

Debloating techniques can be categorized into two main classes: \textit{fixed} and \textit{\DT{}} techniques.
\textit{Fixed debloating techniques} such as SHARD\cite{abubakar2021shard}, Minion\cite{kim2018securing}, and FACE-CHANGE~\cite{gu2014face}
analyze applications throughout the entire lifetime, 
without considering how applications' requirements change during different execution phases. 
For instance, critical firmware code or system calls like \texttt{execve} or \texttt{fork} are usually executed 
only during the initial moments of an application's lifetime. 
However, \textit{fixed debloating techniques} often fail to disable such system calls or code, 
leaving these available to be used in exploits~\cite{ghavamnia2020temporal}. 

In contrast to \textit{fixed debloating techniques}, which restrict specialization granularity to the entire application's execution lifetime, 
\textit{\DT{} debloating techniques}~\cite{ghavamnia2020temporal,rajagopalan2023syspart} 
consider the changing requirements of an application through different execution phases.
For example, Ghavamnia et al.  ~\cite{ghavamnia2020temporal} and Rajagopalan et al.~\cite{rajagopalan2023syspart} proposed  temporal system call specialization techniques for server applications, 
which restrict available system calls based on the application's execution phase.
They use static analysis techniques to identify the required system calls for two distinct 
phases—\textit{initialization} and \textit{serving}—and block unneeded system calls, such as \texttt{fork} and \texttt{execve}, 
after transitioning to the \textit{serving} phase, 
using a Linux-based mechanism called the Seccomp-BPF filter~\cite{Seccomp}.

Unfortunately, \DT{} debloating techniques~\cite{ghavamnia2020temporal,rajagopalan2023syspart} have two limitations. 
First, similar to fixed debloating techniques, they focus on disabling security-critical system calls such as  \texttt{execve}. 
However, embedded devices' firmware, particularly those used in RVs (e.g., ArduPilot~\cite{ardupilot}, PX4~\cite{px4}), 
contains critical functions, such as \texttt{disarm\_motors}, 
which should only be executed in specific modes\footnote{We use \textit{modes} (e.g., \texttt{RTL}, \texttt{AUTO}) to refer to the \textit{phases} of the RV.}, e.g.,
\texttt{Turtle} mode (i.e., flipping the copter). If not, they can lead to serious safety violations.

Second, these techniques use Seccomp-BPF to restrict the set of available system calls for each phase.
However, the Seccomp-BPF mechanism for filtering system calls is irreversible, 
meaning that once a system call is filtered out, it cannot be reallowed later.
In contrast to server applications, which typically have a simple control flow 
and do not revert to the previous phase, 
embedded devices often experience frequent mode changes, 
including transitions back to earlier modes (e.g., \texttt{AUTO <-> LOITER}).
Therefore, debloating techniques that cannot \textit{dynamically} adjust access restrictions on firmware functions at runtime, 
such as those relying on Seccomp-BPF, are unsuitable for embedded devices.


To address the above limitations, we propose \RV{}, an \DT{} debloating approach that automatically identifies 
the set of functionalities required by each mode (e.g., \texttt{TAKEOFF}) at the function-level and then monitors the device at runtime
to ensure that only those firmware functions are allowed to execute when the device operates in the corresponding mode.

\textit{Our key insight is that many embedded devices-such as RVs—operate in distinct modes 
(e.g., \texttt{LAND} or \texttt{TAKEOFF}) at runtime, 
with each mode requiring a specific set of functionalities, implemented as functions in their firmware. We call these devices \emph{mode-based} embedded devices. } 
For instance, when RVs are in \texttt{AUTO} mode, travelling from point A to B, 
they do not require the motor disarming 
function designed for \texttt{LAND} mode; misuse of this function can result in crashes.
\textit{Thus, we can restrict these functionalities (firmware functions) to their respective modes at runtime.}

\RV{} has two main goals: 1) Specializing firmware code by \textit{dynamically} adapting it to changes in the device's requirements 
at the mode level throughout its execution lifetime, 
and 2) Enforcing restrictions on firmware code at \textit{function-level granularity},
providing finer specialization than system call-level~\cite{ghavamnia2020temporal,abubakar2021shard,rajagopalan2023syspart}
or hardware-level (e.g., disabling unused peripherals)~\cite{hu2022irqdebloat,hu2023hacksaw}  techniques. 

\RV{} addresses the above goals through two innovations.  
First, it uses either \textit{static}
or \textit{dynamic analysis} to identify per-mode firmware code coverage, 
while considering changes in the device's requirements throughout its execution
(i.e., the set of firmware functions required for each mode).
For static analysis, \RV{} uses points-to analysis~\cite{sui2016svf,andersen1994program} with two pruning heuristics
to construct a pruned call graph of the entire firmware (e.g., ArduPilot and PX4) and reduce spurious targets for indirect call sites;
it then determines the \textit{reachable (required)} functions for each mode based on the call graph. 
For dynamic analysis, \RV{} employs a dynamic profiling technique by executing the device with different workloads (e.g., missions for RVs) 
to identify the functions 
\textit{required} for each mode at runtime. 
\RV{} performs dynamic profiling by instrumenting firmware functions, 
as well as the mode-switching function responsible for changing device modes 
and tracking mode transitions and function executions.

Second, \RV{} uses a software-based enforcement technique that statically instruments indirect
control flow transfer instructions 
(e.g., indirect call sites and function return instruction) and the mode-switching function
to redirect control flow, monitor function execution and mode transitions, 
and restrict access to functions belonging to other modes. 
Upon a mode switch, 
\RV{} loads the required firmware functions  for the new mode 
by applying per-mode configurations (goal 1), including \textit{required} 
functions (goal 2) identified by either \textit{static} 
(\RVC{}) or \textit{dynamic analysis} (\RVA{}).

We evaluated \RV{} using both techniques, 
and the results (\S\ref{sec:results}) show that \RVA{} can debloat more functions, 
 without introducing either \textit{false negatives} or \textit{false positives} 
(i.e., all allowed functions are used during the missions, and 
no mission fails due to a missing function).
This is because embedded devices typically require a fixed set of functions (functionalities) for each mode, 
which can be identified with just a few missions. 
In contrast, \RVC{} eliminates fewer unneeded firmware functions for each mode, 
while 
employing a more careful and conservative debloating approach.
Thus, the choice between these two approaches involves a trade-off: 
\textit{static analysis} provides more cautious restrictions with lower protection,
whereas \textit{dynamic analysis} offers higher protection but comes with a risk of potential false positives\footnote{Henceforth, \RV{} refers to both variants, unless the variant is explicitly mentioned.}.



\textit{To the best of our knowledge, \RV{} is the first fine-grained, 
\DT{} debloating framework for mode-based embedded devices 
that automatically identifies and specializes the firmware code 
at the function-level for each mode.} 

Our main contributions are summarized as follows:

\begin{itemize}
    \item Propose two automated analysis techniques (static and dynamic) to identify the required firmware code 
    at the function-level, considering the different operational requirements of RVs  
    across different execution modes. 
    \item Propose a runtime \DT{} debloating 
    technique that tracks mode switches and  
    a novel software-only enforcement technique 
    to dynamically specialize the firmware code for each mode at runtime.
    \item Design \RV{}, 
    a fine-grained debloating framework to integrate the above techniques 
    for mode-based embedded devices such as RVs.
    \RV{} is implemented using the LLVM compiler~\cite{llvmcomp},  
    making it portable across different devices. 
    \item Evaluate \RV{} 
    on four simulated and two real RVs, all running one of two widely used open-source autopilot 
    systems, ArduPilot or PX4, across a diverse range of scenarios and missions. We also evaluated it against three realistic code-reuse attacks on the RVs.
\end{itemize}

The results demonstrate that (1) \RVC{} 
reduces the attack surface by an average of 41\% of the 
 firmware functions. 
(2) \RVA{} identifies all \textit{required} functions for each mode 
by running RVs with only up to 10 missions, and achieves an average attack surface reduction of 85\%.
(3) The average performance overhead and power consumption overhead of \RV{} on real RVs are 3.9\% and 0.47\%, respectively.
Further, the overall mission time was not increased due to \RV{}.
(4) The firmware instrumented by \RV{} incurred an average memory overhead of about 4\%, which is about 0.25 MB, and 
(5) \RV{} successfully prevented all three attacks. 

\section{Background}
\label{sec-back}

\subsection{Memory Corruptions}
\label{Sec:Back-memory}
The majority of embedded devices' software, including firmware, is developed in low-level programming languages 
(i.e., C or C++) due to their efficiency and ability to provide full control over the underlying hardware. 
However, these languages are type-unsafe, requiring developers
to carefully manage memory accesses and ensure their validity to prevent issues like dereferencing invalid pointers. 
The lack of memory safety in these languages, combined with developers' failure to meet these responsibilities,
leads to memory corruption vulnerabilities (e.g., buffer overflow) that attackers can exploit 
to manipulate device behavior maliciously or even gain full control over 
the system's execution flow~\cite{szekeres2013sok} (e.g., control flow hijacking attacks).
Control flow hijacking attacks occur when a code pointer, 
such as a return address (backward edge) or a function pointer (forward edge),
on the backward edge (e.g., return addresses) or forward edge (e.g., function pointers) 
becomes corrupted. 
By hijacking the control flow, attackers can either execute malicious payloads injected
into the application stack (\textit{code injection attacks}) or reuse existing code sequences (gadgets) 
from the victim application (\textit{code reuse attacks}). 
Security mechanisms like Data Execution Prevention (DEP), 
which mark memory regions as either writable or executable (i.e., $W \oplus X$), mitigate code injection attacks. 
However, code reuse attacks bypass these defenses by injecting addresses of existing instructions 
into corrupted code pointers, altering the original control flow in an arbitrarily expressive way. 

\subsection{Mode-based Embedded Devices}
\label{sub-sec:mode-based}
\begin{figure}[h]
    \centering
    \begin{subfigure}[b]{0.38\columnwidth}
        \centering
        \includegraphics[scale=0.18]{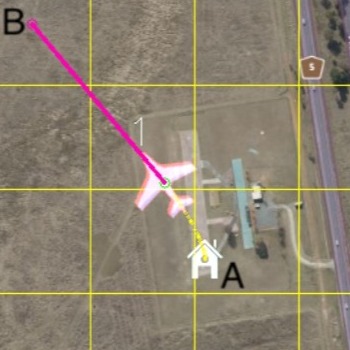}
        \caption{ArduPlane}
        \label{fig:mission1}
    \end{subfigure}
    \hspace{0.01\columnwidth}
    \begin{subfigure}[b]{0.38\columnwidth}
        \centering
        \includegraphics[scale=0.18]{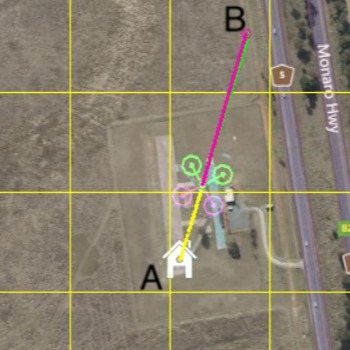}
        \caption{ArduCopter}
        \label{fig:mission2}
    \end{subfigure}
    \caption{Two mission examples for ArduPilot firmware.}
    \label{fig:missions}
\end{figure}

Embedded devices are widely used in mission-critical applications such as 
smart homes, industrial control systems, and autonomous vehicles.
One class of embedded devices are mode-based devices, which operate in different modes during their tasks (missions).
For instance, robotic vehicles (RVs) switch between different modes such as
\texttt{STABILIZE}, \texttt{AUTO}, and \texttt{RTL} while performing missions like delivering a package from point A to B.
Examples of simple missions for the ArduPilot firmware are shown in Figure~\ref{fig:mission1} and Figure~\ref{fig:mission2} for ArduPlane and ArduCopter, respectively.

\begin{figure}[h]
    \begin{subfigure}[t]{\columnwidth}
        \centering
        \includegraphics[scale=0.39]{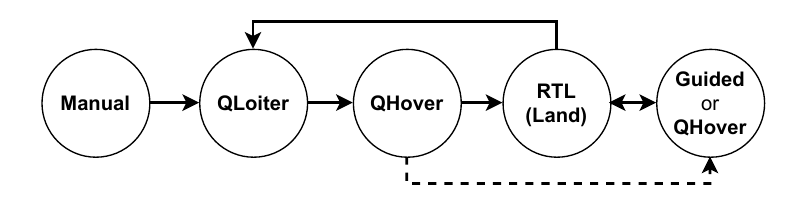}
        \caption{Mode switching process in ArduPlane.}
        \label{subfig:RAVModes}
    \end{subfigure}\hfill
    
    \begin{subfigure}[t]{\columnwidth}
        \centering
        \includegraphics[scale=0.39]{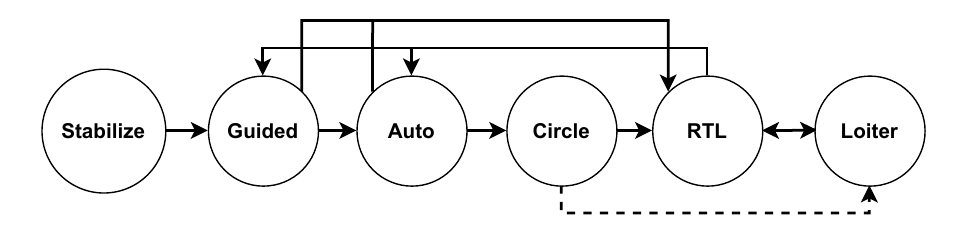}
        \caption{Mode switching process in ArduCopter.}
        \label{subfig:RAVModesCopter}
    \end{subfigure}
    \caption{Mode switching for ArduPlane and ArduCopter.}
    \label{fig:RAV-modes}
\end{figure}

These modes are predefined in RVs' autopilot software such as ArduPilot~\cite{ardupilot} and PX4~\cite{px4}.
Further, to manage their tasks, mode-based embedded devices often rely on real-time operating systems (RTOS) such as NuttX~\cite{NuttXcite}.
All these software components—the autopilot software, RTOS, and other essential modules like device drivers—are integrated into the device's firmware.
As an example, Figure~\ref{subfig:RAVModes} and Figure~\ref{subfig:RAVModesCopter} present mode switching 
in \textit{ArduPlane}~\cite{ArduPDoc} and \textit{ArduCopter}~\cite{ArduCDoc}, sub-platforms of the ArduPilot software.
The list of ArduPilot and PX4 modes is presented in Table~\ref{table:modeexplanation} in the Appendix. 


Many embedded devices, including mode-based ones, often lack hardware security features, such as Memory Management Unit (MMU), 
which results in all applications being mapped into a single memory space~\cite{kim2018securing}.
Therefore, these applications have unrestricted access to the entire shared memory in the privileged mode,
enabling attackers to exploit a single memory corruption vulnerability 
to manipulate the control flow of applications and take over the entire system. 

Furthermore, these devices may not support Memory Protection Units (MPUs)~\cite{salehi2020musbs}. 
Even when an MPU is available, configuring and controlling memory access presents challenges 
due to its limitations. Unlike the MMU, 
which enforces access restrictions through page tables and virtual memory, 
an MPU operates with a limited number of hardware registers, 
typically supporting only 8 to 16 memory regions (e.g., 8 regions in ARMv7-M~\cite{clements2018aces}). 
Each region is protected using read, write, and execute permissions, 
configurable for both privileged and unprivileged modes. 
Also, MPU regions must be a power of two in size, with a minimum size of 32 bytes, 
and their starting addresses must be a multiple of their size. This severely restricts MPUs' use. 


\label{backsec}

\section{Motivation}
\label{sec-motivation}

\subsection{Debloating Techniques' Limitations}
\label{sub-sec:current-limitations}
    \textbf{L1: Treating Applications as Monolithic Entities with Fixed Requirements.}
    \textit{Fixed techniques} in attack surface reduction~\cite{qian2019razor,heo2018effective,abubakar2021shard,kim2018securing}
    specialize or remove functionality based on the entire application's lifetime requirements.
    However, \textit{\DT{} debloating techniques}~\cite{ghavamnia2020temporal,rajagopalan2023syspart} 
    have shown that considering the changing requirements during the application's execution phases is more effective.
    In particular, many devices and programs, such as mode-based embedded devices and server applications, 
    operate in distinct modes, with their requirements changing based on the execution phase. 


    \textbf{L2: Removing Functionalities.}
    \textit{Fixed debloating techniques}~\cite{hu2022irqdebloat,hu2023hacksaw} often remove functionalities, such as hardware features or kernel source code, 
    that are not required at any point during the applications' lifetime. 
    On the other hand, these techniques retain critical functionalities used for legitimate purposes during the application's operation, 
    while some of these functionalities are only executed for a short period of time.
    Therefore, to prevent over-debloating and potential system crashes, these techniques often overestimate the functionalities required.
    Retaining such functionalities would inflate the attack surface. 

    \textbf{L3: Using MPU.}
    Some \textit{Fixed debloating techniques}~\cite{kim2018securing,clements2018aces} use MPUs to control memory access. 
    However, this hardware may not be available on all devices, 
    particularly embedded systems~\cite{salehi2020musbs}. Moreover, due to MPU limitations (\S\ref{sub-sec:mode-based}), 
    these methods must rely on clustering techniques to group code for multiple processes. 
    Thus, processes may gain access to unrequired code, thus leading to security risks.

    \textbf{L4: Considering System Calls.}
    Existing debloating techniques, both \textit{fixed} and \textit{\DT{}}, only focus on system calls to reduce attack surfaces.
    However, other functionalities implemented as firmware functions, such as those in RV autopilot software, 
    may also pose security risks if accessible throughout the device's execution lifetime.
    For instance, Listing~\ref{list1} and Listing~\ref{list} show two security-critical functions 
    in ArduPilot for the \texttt{Land} and \texttt{Turtle} modes, respectively.
    The purpose of the function shown in Listing~\ref{list1} is to land the RV when it is in 
    specific states, such as upon reaching its destination.
    Also, the function in Listing~\ref{list} is executed to disarm the motors if
    the RV lands upside down, and enters the \texttt{Turtle} mode to flip the drone upright.
    However, executing \texttt{disarm\_motors} during flight immediately kills the motors, causing a crash.
    Thus, these functions should not be accessible in other modes, 
    such as \texttt{AUTO} or \texttt{GUIDED}, as attackers could exploit vulnerabilities to execute these functions
    and manipulate the device's behavior.



    \textbf{L5: Irreversible Mechanism.}
    Although \textit{\DT{} debloating techniques} like SysPart~\cite{rajagopalan2023syspart} and TSCP~\cite{ghavamnia2020temporal}
    are more effective at reducing attack surfaces compared to \textit{fixed techniques},
    they employ an irreversible mechanism to disable system calls after transitioning to the \textit{serving} phase.
    Specifically, these techniques leverage a Linux kernel feature called Seccomp-BPF, which is a one-way process,
    meaning the filtered system calls cannot be re-enabled. 
    As a result, they are suitable for applications with simple control flow, such as \textit{server programs}, 
    but not for devices like \textit{mode-based embedded devices} that frequently switch between modes,  and 
    transition back to earlier modes (Figure~\ref{fig:RAV-modes}). 
    
\begin{lstlisting}[caption=Landing function for ArduCopter's Land mode.,label=list1]
1 void ModeLand::run () {
2   if (control_position){
3        gps_run();
4    }else {
5        nogps_run();
6    }	
7 }
\end{lstlisting}
    
\begin{lstlisting}[caption=ArduCopter's function executed in Turtle mode to disarm the vehicle motors if it lands upside down.,label=list]
1 void ModeTurtle::disarm_motors(){
2    //... omit
3    motors->armed(false); // disarm
4     //... omit
5    hal.util->set_soft_armed(false); 
6 }
\end{lstlisting}



\subsection{Our Approach: \RV{} }
\label{sub-sec:solution}

\RV{} proposes two solutions (i.e., S1 and S2) 
to address the limitations of existing debloating techniques.

    \textit{\textbf{S1: Addressing L1 and L4.}}
    We propose an \DT{} debloating technique that specializes firmware code 
    based on the device's requirements during its mode of execution (L1).
    This technique, detailed in \S\ref{sub-sec:offline-analysis}, 
    employs either a static or dynamic analysis 
    to identify the required firmware code for each mode at the function-level (L4).

    \textit{\textbf{S2: Addressing L2, L3, and L5.}}
    We use a software-based technique that does not rely on hardware features
    to enforce firmware debloating 
    based on the device's current execution mode throughout its execution lifetime (L5).
    Specifically, \RV{} monitors mode switches and controls memory access by instrumenting the firmware's mode-switching functions 
    and indirect control flow transfer instructions to restrict access to required firmware functions based on the device's current mode (L2 and L3).


\subsection{Assumptions and Threat Model}
\label{subsec-assumption}
\textbf{Assumptions.}
We make the following two assumptions.
First, the list of autopilot software modes (e.g., \texttt{TAKEOFF} and \texttt{AUTO}) 
and mode-switching functions responsible for changing modes at runtime are known to \RV{} and provided as inputs to it. 
This information can be found in the firmware's source code or autopilot software's documentation (e.g., ArduPilot documentation~\cite{ArduPDoc,ArduCDoc}).
Although determining mode-switching function could be automated through dynamic analysis, 
we did not invest the effort to develop this capability,
as it only needs to be performed once per firmware model (e.g., ArduPlane). 
In our experiments, it took us only a few minutes to identify this function in six different firmware models (e.g., ArduPlane and PXCopter).

Second, similar to prior debloating techniques~\cite{abubakar2021shard,ghavamnia2020temporal,kim2018securing}, we assume that \RV{} has access to the firmware's source code. 
With that said, since \RV{} operates on the LLVM IR~\cite{llvmcomp}, it can also work if the firmware is available in the LLVM Intermediate Representation (IR) format instead of source code. 

\textbf{Threat Model.}
We consider an adversary who is capable of performing remote attacks on the system without having root or physical access to it. 
Furthermore, we assume that the core system software is benign and developed by honest developers; however it may contain
memory corruption vulnerabilities (e.g., buffer overflows). 
An adversary can exploit 
such memory corruption vulnerabilities,
to hijack the firmware's control flow and call unneeded critical firmware functions via code-reuse attacks. 
Our threat model is consistent with those of prior debloating techniques~\cite{abubakar2021shard,kim2018securing}.



\label{motivsec}

\section{Design: \RV{}}

This section presents the system design of \RV{}, the first \DT{} debloating technique for \textit{mode-based embedded devices}. 
We start with an overview of \RV{} and its key innovations (\S\ref{sub-sec:overview}), before 
describing the different phases of the system in detail (\S\ref{sub-sec:offline-analysis}, \S\ref{sub-sec:offline-instrumentation-monitoring} and \S\ref{sub-sec:runtime-monitor}).

\subsection{Overview of \RV{}}
\label{sub-sec:overview}

\begin{figure*}[ht]    
	\centering
	\includegraphics[scale=0.56]{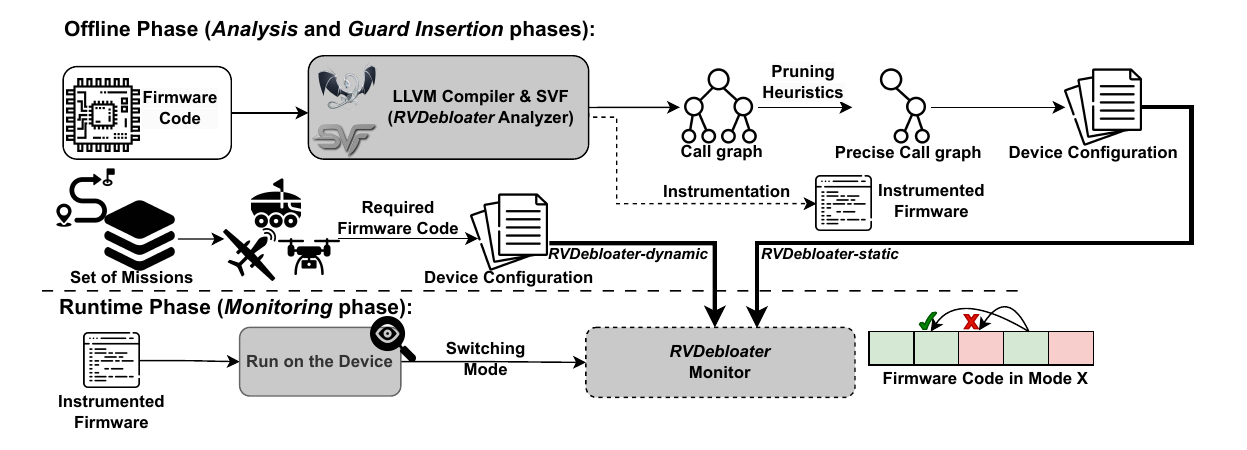} 
	\caption{Overview of \RV{}'s operation, including both the offline phase (top) and runtime phase (bottom). }
	\label{OverviewD}
\end{figure*}

Figure~\ref{OverviewD} shows \RV{}'s workflow, which includes three main phases: 
(1) the analysis phase, (2) the guard insertion phase, and (3) the monitoring phase.
Our approach aims to identify and dynamically specialize firmware code\footnote{Mode-based embedded device firmware consists of code from various software modules, such as libraries and applications, compiled into a single executable file.} 
running on mode-based embedded devices 
at the function-level granularity (i.e., firmware functions), 
depending on the device's execution mode.
To accomplish this goal, we propose two key innovations.
The first innovation is using either a static analysis 
or dynamic analysis
to identify the firmware functions used in a specific mode (see S1 in \S\ref{sub-sec:solution}). 
We leverage this method in the \textit{analysis} phase.

The second main innovation of our technique is that, 
unlike previous debloating methods that rely on hardware components (MPU)~\cite{kim2018securing}, 
virtualization mechanisms (VMX)~\cite{abubakar2021shard}, or irreversible mechanisms (Seccomp-BPF)~\cite{ghavamnia2020temporal,rajagopalan2023syspart}, 
our approach uses a completely \textit{software-based enforcement technique} (see S2 in \S\ref{sub-sec:solution}).
In particular, our \textit{software-based enforcement technique} 
uses \textit{static analysis} to instrument indirect control flow transfer instructions (e.g., indirect function
call or function return instruction) and redirect the control flow to the \RV{} monitor,
which specializes firmware code for each mode using a mode-specific configuration file 
that includes the required functions and their addresses generated in the previous phase (i.e., \textit{analysis} phase). 
This is achieved in two phases:
(1) the \textit{guard insertion} phase, which modifies the firmware code, 
and (2) the \textit{monitoring} phase, which tracks the device's execution to check indirect control flow transfer instructions and specialize the firmware for each mode.
As shown in Figure~\ref{OverviewD}, the \textit{analysis} and \textit{guard insertion} phase are both performed offline, 
while the \textit{monitoring} phase is performed online.

After generating the mode-based configuration file (\textit{analysis} phase) and executing the instrumented firmware (\textit{guard insertion} phase) on the device, 
the \RV{} monitor is triggered when an indirect control flow transfer instruction is executed (\textit{monitoring} phase). 
The monitor checks whether the jump and execution of the target  address are allowed based on the configuration file. 
If the target is not present in the static output (\RVC{}) or
in the dynamic output (\RVA{}) (depending on which is used), the monitor can either
activate safety mechanisms such as \texttt{Fail-Safe} mode
or take other actions
based on the device maintainer's preferences (e.g., mechanisms they have developed). 
\texttt{Fail-Safe} mechanisms are commonly employed in various detection 
and recovery techniques~\cite{li2025software,choi2018detecting,kim2018securing,quinonez2020savior}
to safely control the vehicle by returning it to the launch point or 
performing an emergency landing
when an attack is detected or control is lost.
ArduPilot provides 11 well-defined fail-safe mechanisms~\cite{ardufail}, and PX4 also includes a 
fail-safe landing~\cite{px4fail}. Thus, we used fail-safe mechanisms in our evaluation.

We provide a detailed discussion of the process in the following sections (\S\ref{sub-sec:offline-analysis}, \S\ref{sub-sec:offline-instrumentation-monitoring} and \S\ref{sub-sec:runtime-monitor}).
Note that, for simplicity, the examples are presented in the C language, 
even though \RV{} operates on LLVM IR. 
An example of LLVM IR is provided in \S\ref{sec-app:llvmir}.
Further, to describe the examples, 
we treat each line of C code (i.e., program statement) as a single instruction, even though, 
in practice, each of these would correspond to multiple LLVM instructions.


\subsection{Analysis Phase}
\label{sub-sec:offline-analysis}
The first phase of the \RV{} process is the offline analysis phase, 
which automatically identifies the \textit{required functions} for each mode using either \textit{static} or \textit{dynamic} analysis. 
Since static analysis tends to overestimate the required functions for each mode, 
while dynamic analysis tends to underestimate them, \RV{} leverages \textit{both techniques},
giving device maintainers the flexibility to choose the most appropriate option based on their level of risk-tolerance.

Specifically, the offline analysis phase consists of either: 
(1) executing the device with different missions to dynamically identify the functions 
used in each mode at runtime (\RVA{} \S\ref{subsub-section:dynamic-analysis}), 
OR (2) generating a pruned call graph of the firmware to statically identify an overestimated 
set of required functions for each mode (\RVC{} \S\ref{subsub-section:static-analysis}). 
%
Subsequently, \RVA{} uses the dynamic output
to enforce firmware specialization based on the device's current mode,
while \RVC{} uses the identified function lists for each mode from the static output.
In the following sections, we explain each step in the offline analysis phase.

\subsubsection{\textbf{\RVA{}: Dynamic Analysis}}
\label{subsub-section:dynamic-analysis}
To enforce debloating, \RV{} leverages dynamic analysis to identify the required firmware functions 
for each mode (\RVA{}). To do so, \RV{} employs static instrumentation to generate a modified firmware version 
that records (1) mode switches, and (2) the firmware functions executed in each mode at runtime.
Afterwards, \RV{} runs the instrumented firmware on the device with various missions 
to determine the executed (\textit{required}) firmware functions for each mode. The following outlines each step of the dynamic analysis in detail.

\textbf{(1) Instrumentation.}
This step inserts trampolines\footnote{A single instruction, like a jump or branch instruction, inserted into the program to redirect its original control flow.} 
into the mode-switching function(s)\footnote{Although \RV{} supports multiple mode-switching functions, in the tested firmware (e.g., ArduCopter, PXCopter), instrumenting a single mode-switching function was sufficient to monitor all mode changes.} and firmware functions.
\RV{} provides an instrumentation component that automates this process.
This component takes the firmware and mode-switching function(s) (\S\ref{subsec-assumption}) as input and
outputs the instrumented firmware.
Specifically, the instrumentation component statically analyzes the firmware to locate the given mode-switching function(s) 
and other firmware functions,
then inserts the trampolines.
To ensure that the mode has changed successfully, the component inserts a trampoline at the end of mode-switching function(s). 
If a mode switch fails, the firmware handles the error independently of \RV{}. 

\begin{lstlisting}[caption=Output of the instrumentation step for the mode-switching function in ArduPlane (ArduPilot).,label=instrumentation_example]
1 bool Plane::set_mode_by_number(const Mode::Number new_mode_number, const ModeReason reason){
2    Mode *new_mode = plane.mode_from_mode_num(new_mode_number);
3    if (new_mode == nullptr) {
4        notify_no_such_mode(new_mode_number);
5        return false;
6    }
7    %*\colorbox{greenannoback}{Trampoline}*)
8    return set_mode(*new_mode, reason); }
\end{lstlisting}


Listing~\ref{instrumentation_example} shows an example of the mode-switching function in ArduPlane. 
This function first (line 2 in Listing~\ref{instrumentation_example}) retrieves the name of the new mode (e.g., \texttt{CIRCLE} or \texttt{RTL}) 
by passing the input number (\texttt{new\_mode\_number}).
Then, it checks the returned mode name to validate whether the new mode exists (lines 3 to 6). 
Finally, it calls a function to switch the current mode to the new mode (line 8).
As seen in Listing~\ref{instrumentation_example}, the component has inserted a trampoline at the end of the function (line 7)
after the mode change has been successfully validated (lines 3 to 6).
When the inserted trampoline is triggered at runtime, it passes the current and new mode to \RV{}'s profiler, 
records the executed functions in the current mode, and changes the profiler's mode to the new mode. 

	

Further, the component inserts trampolines at the beginning of all firmware functions (except the mode-switching function) 
to ensure that every function executed in the current mode triggers \RV{}'s profiler. 
Algorithm~\ref{alg_ins} demonstrates the workflow of the instrumentation step. 
For the mode-switching function, the component inserts a trampoline at the end (lines 2 to 6). For all other functions, it adds trampolines to the beginning (lines 7 to 11).
\begin{algorithm}[h]
\caption{Instrumentation Component for Profiling}
\label{alg_ins}
\begin{footnotesize}
\begin{algorithmic}[1]  

\algrenewcommand\alglinenumber[1]{}
\Input Firmware functions, mode-switching function set $modeFuncs$
\Output Instrumented firmware

\setcounter{ALG@line}{0}  
\algrenewcommand\alglinenumber[1]{\arabic{ALG@line}\hspace*{3mm}}

\Function{InstrumentationComponent}{$Functions$}

    \If{$Function \in modeFuncs$}
        \If{$End$ $of$ $Function$}
            \State $location \gets findLocation()$
            \State insertTrampoline($location$)
        \EndIf
    \Else   \Comment{Other firmware functions}
        \If{$Function$ $Entrance$}
            \State $location \gets findLocation()$
            \State insertTrampoline($location$)
        \EndIf
    \EndIf

    \State \Return $instrumentedFirmware$

\EndFunction

\end{algorithmic}
\end{footnotesize}
\end{algorithm}


\textbf{(2) Dynamic Profiling.}
In the next step, \RV{} runs the device with the instrumented firmware (e.g., ArduPlane or ArduCopter) across different missions 
and scenarios (e.g., with and without obstacles) to identify the firmware functions for each mode at runtime
by capturing functions with \RV{}'s profiler. 
The instrumented firmware ensures that each function for the current mode is logged at most once when the device switches to a new mode.
We assume that \RV{} runs this step (dynamic profiling) in a benign environment, free from any security attacks.

\subsubsection{\textbf{\RVC{}: Static Analysis}}
\label{subsub-section:static-analysis}

\RV{} also statically analyzes the firmware to construct its call graph. 
Since the firmware is written in C/C++, it often relies on indirect function calls via function pointers. 
Statically resolving the targets of these calls is both challenging and critical 
for accurately identifying the functions required by each mode. 
To address this, \RV{} employs Andersen's points-to analysis~\cite{andersen1994program}, 
a static technique for resolving pointer targets in C/C++  programs. 
This analysis can have different sensitivities depending on how objects in memory are modeled:
field sensitivity, path sensitivity, and context sensitivity.
Higher levels of sensitivity yield more accurate results but also increase the analysis time 
and the implementation complexity.
For instance, field sensitivity distinguishes and models each field of a structured variable (e.g., a \texttt{struct}) separately,
while context sensitivity distinguishes between different call sites of a function, 
meaning that when the same function is called from different places, each call gets 
its own context and is analyzed independently of the others.
While a detailed discussion of these topics is beyond the scope of this paper, 
more details about these sensitivities 
can be found in previous work~\cite{ghavamnia2020temporal,hind2001pointer}.
Specifically, \RV{} leverages SVF~\cite{sui2016svf}, which implements Andersen's analysis with field sensitivity but without path or context sensitivity. 

As a result, SVF tends to over-approximate the call graph, 
as it associates indirect function calls with a large number of potential targets—only a small subset of 
which is actually executed, while the rest are spurious. 
Although this over-approximation increases the number of functions identified per mode, 
introducing unneeded functions that increase the attack surface and could be exploited by attackers to subvert the system,
the approach is sound in determining which functions are not required (unreachable) for each mode.
However, to improve precision, \RV{} applies two pruning heuristics, 
inspired by the TSCP technique~\cite{ghavamnia2020temporal}, 
to remove spurious targets caused by the lack of path and context sensitivity
and generate a more accurate call graph. 
We modified these heuristics to support embedded firmware 
and refine call graphs further. 
The call graph is generated once per firmware model (e.g., ArduCopter), 
independent of the number of modes.
The two heuristics are as follows. 

\textbf{(1) Signature-based Pruning Heuristic.}
SVF resolves indirect call sites using a basic pruning heuristic that considers only the number of arguments. 
However, this approach does not effectively prune spurious targets, 
as it considers functions with the correct number of arguments but incorrect argument types 
as valid targets. 

To improve precision, we additionally use 
two filtering methods.
First, \RV{} considers the 
argument types when resolving indirect call targets, 
discarding those that do not match. 
In particular, \RV{} analyzes all indirect call sites and prunes call edges to functions whose argument types do not match the expected types at the call site.
Second, \RV{} enhances call graph accuracy by also considering return types. 
Specifically, functions with a \texttt{void} return type are identified as spurious targets 
for indirect call sites that expect a return value.
It effectively prunes spurious targets and reduces the number of call edges in the generated call graph. 

\textbf{(2) Address-based Pruning Heuristic.}
To further prune spurious targets, \RV{} applies an additional heuristic (address-based pruning heuristic) based on ``taken"  addresses in the reachable path to the entry point.  
Andersen's algorithm assumes that \emph{all functions in the program} are reachable from the entry point. 
However, a function may not have its address stored in a variable or may be unreachable through the entry point, 
making it an invalid target for an indirect call site. 
 
Thus, \RV{} identifies all functions whose addresses appear in paths reachable from the entry point. 
Then, it examines all indirect call sites and removes edges that point to functions that are not in this list—i.e., 
functions whose addresses are not in any reachable path from the entry point.

After refining the call graph, \RV{} identifies the entry functions of each mode to 
determine their reachable (required) functions. 
To do so, it employs an instrumentation component similar to the one used in dynamic analysis (\S\ref{subsub-section:dynamic-analysis}), 
which instruments both the firmware functions and the mode-switching function(s).
\RV{} then executes the device through a mission that includes all modes intended for monitoring and control at runtime by \RVC{}.   
During execution, when the device switches modes, 
\RV{} tracks the function(s) invoked by the mode-switching function to initialize the parameters 
required for the new mode and marks them as the entry function(s) of that mode. 

\RV{} also performs mode-name pattern matching to confirm the validity of the identified entry function(s). 
In particular, we observed that these firmware models (e.g., ArduCopter or ArduPlane) have entry function(s) 
responsible for managing mode-specific parameters, and their names typically include the mode name itself.
Therefore, \RV{} uses the list of mode names provided as input (\S\ref{subsec-assumption})
to verify and confirm the correctness of these functions.
We found that all identified entry functions were correct. 
Note that this pattern can be adapted by developers to support other firmware. 
However, we used the same pattern in all the firmware models evaluated in this paper, as it was sufficient.
Finally, \RV{} uses the generated call graph to 
identify the reachable and unreachable functions for each mode, starting from its entry function(s).

Both \textit{dynamic} and \textit{static} analyses only need to be performed \textit{once} for each firmware model
(e.g., ArduPlane or ArduCopter) to identify and store the functions required for each mode at runtime. 
These identified functions are then used during the runtime (device execution) monitoring phase to enforce debloating 
by checking whether they are permitted in the current mode of the device.

\subsection{Guard Insertion Phase}
\label{sub-sec:offline-instrumentation-monitoring}

The next phase of the \RV{} process is the \textit{guard insertion} phase, which is performed offline and only once per firmware. 
The \RV{} \textit{monitoring} phase is designed to: (1) track mode switches, 
(2) load the appropriate list of identified functions for the new mode when a mode switch occurs,
and (3) monitor indirect control flow transfer instructions and ensure they are allowed; otherwise, \RV{} logs the violation and switches to \texttt{Fail-Safe} mode.
To achieve these goals, the \RV{} process has two main phases: 
\textbf{guard insertion} and \textbf{monitoring} (\S\ref{sub-sec:runtime-monitor}).

Similar to the previous instrumentation step for offline analysis phase, 
\RV{} provides a component that automatically identifies mode-switching function(s) and inserts a trampoline at the end of each.
However, in this instrumentation step, \RV{} 
replaces \emph{all} indirect control flow transfer instructions in the firmware code  
with calls to the \RV{} monitor. 
This allows \RV{} to observe all function calls and returns, and verify them against the	 list of allowed functions identified in the previous phase. 
\begin{algorithm}[h]
	\caption{Instrumentation Component for Monitoring}
	\label{alg_ins_profiling}
	\begin{footnotesize}
	\begin{algorithmic}[1]  
	
	\algrenewcommand\alglinenumber[1]{}
	\Input Firmware functions ($Insts = (inst_1, \dots, inst_n)$) and mode-switching function set ($modeFuncs$)
	\Output Instrumented firmware
	
	\setcounter{ALG@line}{0}  
	\algrenewcommand\alglinenumber[1]{\arabic{ALG@line}\hspace*{3mm}}
	
	\Function{InstrumentationComponent}{$Functions$}
	
		\If{$Function \in modeFuncs$}
			\If{$End$ $of$ $Function$}
				\State $location \gets findLocation()$
				\State insertTrampoline($location$)
			\EndIf
		\Else
			\ForAll{$instruction$ $inst_i \in Insts$}
				\If{$inst_i.\text{type}$ is indirect\_control\_flow\_instruction}
					\State $location \gets findLocation()$
					\State replaceInst($location$)
				\ElsIf{$inst_i.\text{type}$ is return\_instruction}
					\State $location \gets findLocation()$
					\State replaceInst($location$)
				\Else
					\State \textbf{continue}
				\EndIf
			\EndFor
		\EndIf
	
		\State \Return $instrumentedFirmware$
	
	\EndFunction
	
	\end{algorithmic}
	\end{footnotesize}
	\end{algorithm}
	
This component takes a list of mode-switching function(s) and the firmware code as input (\S\ref{subsec-assumption}), 
and outputs an instrumented firmware where trampolines are statically added to the mode-switching functions, and indirect control flow transfer instructions are replaced accordingly.
\textit{As noted in the offline analysis phase (\S\ref{subsub-section:dynamic-analysis}), \RV{} can instrument multiple mode-switching functions. 
However, in the tested firmware in this paper (i.e., ArduPilot, PX4), a single mode-switching function  
handled all the mode changes.}

Algorithm~\ref{alg_ins_profiling} describes the workflow of this instrumentation component for monitoring.
Similar to the prior instrumentation component 
(Algorithm~\ref{alg_ins}), 
this component inserts trampolines at the end of mode-switching functions (lines 2 to 6).
For other firmware functions, it iterates over their instructions, and replaces each 
indirect control flow transfer instruction (e.g., indirect call site or return instruction) 
with an instruction (i.e., \texttt{call "monitor\_fn"}) that redirects the control to 
\RV{}'s monitor (lines 7 to 18). \RV{}'s monitor checks the call target 
before calling the intended target address directly.
Section~\ref{sub-sec:runtime-monitor} provides details.

\subsection{Monitoring Phase}
\label{sub-sec:runtime-monitor}

\begin{figure}[h]    
	\centering
	\includegraphics[scale=0.44]{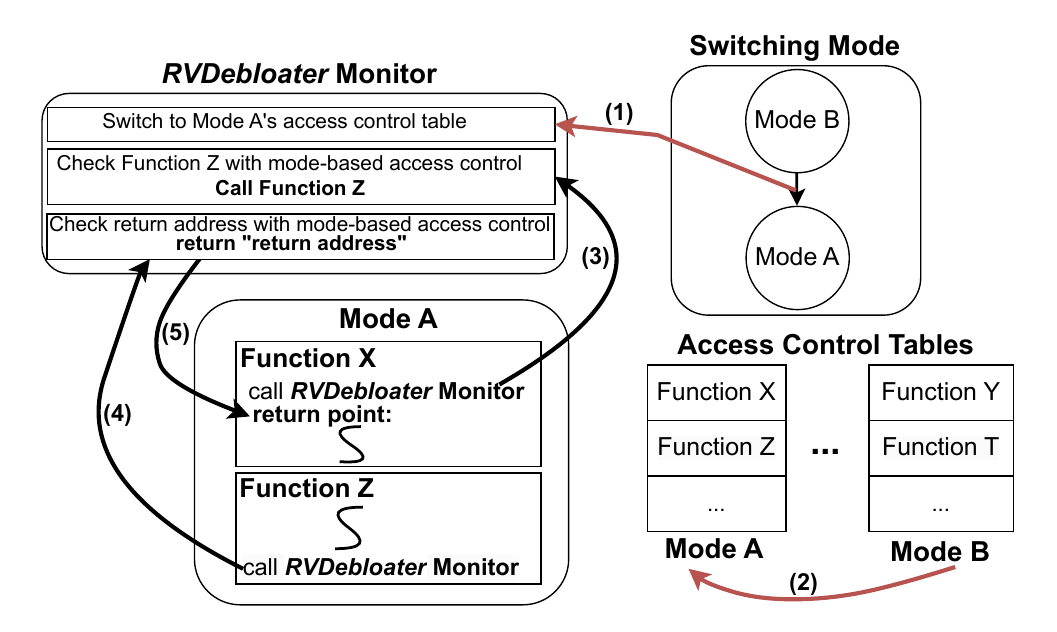}
	\caption{\RV{} Runtime System. Steps (1), (2) occur during a mode change to load the access control table, including functions required for the new mode.  
	Steps (3)-(5) monitor and control indirect control flow transfer instructions based on the current mode's access control table.}
	\label{RuntimeS}
\end{figure}

The final phase of the \RV{} process is the \textit{monitoring phase}, 
which operates after the \textit{analysis} and \textit{guard insertion} phases
have determined the required functions for each mode (either static or dynamic output)
and instrumented the firmware.
The purpose of the monitoring phase is to \textit{monitor} the device and \textit{dynamically debloat} (specialize) the firmware functions for each mode 
based on the identified lists at runtime, such as during RV missions.
To do so, \RV{} uses the instrumented firmware from the previous phase (\S\ref{sub-sec:offline-instrumentation-monitoring}) and
the list of functions identified for each mode, which is generated during the analysis phase (\S\ref{subsub-section:static-analysis} or \S\ref{subsub-section:dynamic-analysis}).
The procedure of monitoring call sites and returns is illustrated in Figure~\ref{RuntimeS}. 
The figure consists of five steps (labelled (1) to (5)) and four phases: 
the process of switching modes in the RV, an access control table for each mode, 
an example of Mode A at runtime, and the \RV{} monitor component.

Once a mode switch occurs, the mode-switching function jumps to \RV{}'s monitor (step (1)) to update the access control tables for the new mode (step (2)).
For example, as shown in Figure~\ref{RuntimeS} (steps (1) and (2)), when the device mode changes from \textit{Mode B} to \textit{Mode A}, the access control table containing the required functions for \textit{Mode A} is loaded from the generated configuration file.

As mentioned, \RV{} rewrites call instructions to redirect them to its monitor 
(step (3) in Figure~\ref{RuntimeS}). 
When Function X calls Function Z, control is transferred to the \RV{} monitor, 
which checks whether the call is permitted using the loaded access control table 
for the current mode (based on either the static or dynamic output).


If calling Function Z is permitted, the monitor saves the return address (e.g., the "return point" in Function X, as shown in Figure~\ref{RuntimeS}) and then invokes the function. 
Otherwise, 
if the function call is not authorized based on the static output for \RVC{} or the dynamic output for \RVA{}, 
\RV{} triggers an alert and switches the device to \texttt{Fail-Safe} mode, as the call is unauthorized according to the respective analysis.

After finishing the execution of Function Z, the return instruction jumps back to the monitor (step (4) in Figure~\ref{RuntimeS}), 
which verifies the stored return address. 
If the address matches the expected return address, control flow returns to Function X (step (5) in Figure~\ref{RuntimeS}). 
Otherwise, the monitor performs the same approach as for an unauthorized function call (i.e., switching to \texttt{Fail-Safe} mode).


\label{designsec}

\section{Implementation}
\label{sec:implementation}



We implemented \RV{}'s analysis phase using two LLVM module passes, 
which can be executed on firmware containing different functions~\cite{llvmpass}. 
Furthermore, we developed two functions  
that are linked to the firmware to support the instrumentation component and \RV{}'s profiler. 
As mentioned in \S\ref{sub-sec:offline-analysis}, this phase takes the mode-switching function(s) 
and firmware code as input and generates an instrumented firmware 
along with a list of required functions for each mode (configuration file). 
We also utilized SVF~\cite{svftool} to generate the firmware's call graph. 
However, we modified and added two LLVM passes to refine the generated call graph 
(i.e., signature-based and address-based pruning heuristics).

Further, the guard insertion phase is implemented as two LLVM module passes, and is executed on the firmware. 
This generates an instrumented firmware that tracks mode changes 
and controls function execution based on the identified required functions. 
Finally, the runtime monitoring phase  
monitors mode switching and indirect control flow transfer instructions to \textit{dynamically} specialize the firmware code for each mode according to the configuration file. 
The implementation details are discussed in \S\ref{seca:implementation-details} in the Appendix.

%

\label{implementsec}

\section{Evaluation}
This section describes the evaluation of different aspects of \RV{} including the effectiveness and efficiency
of the system. Effectiveness is the ability of \RV{} to specialize the firmware code for each mode at runtime and reduce the attack surface,
while the efficiency is the runtime and memory overheads of \RV{}. 

\subsection{Experimental Setup}
\label{subsec:experimental-setup}
We evaluated \RV{} on four simulated RV systems using either ArduPilot~\cite{ardupilot} or PX4~\cite{px4}—two popular 
open-source autopilot platforms—to test whether \RV{} can work across different platforms without modification: 
(1) ArduPilot's quadcopter~\cite{ardupilotcopter} (ArduCopter), (2) ArduPilot's quadplane~\cite{ardupilotplane} (ArduPlane), 
(3) ArduPilot's ground rover~\cite{ardupilotrover} (ArduRover), and (4) PX4's quadcopter~\cite{px4quadrotor} (PXCopter).  
We used Gazebo~\cite{gazebo}, QGroundControl~\cite{qgroundcontrol}, 
and APM SITL (Software-In-The-Loop)~\cite{ardupilot} platforms for simulations.


\begin{wrapfigure}{r}{0.43\columnwidth}
    \centering
    \includegraphics[width=0.19\columnwidth]{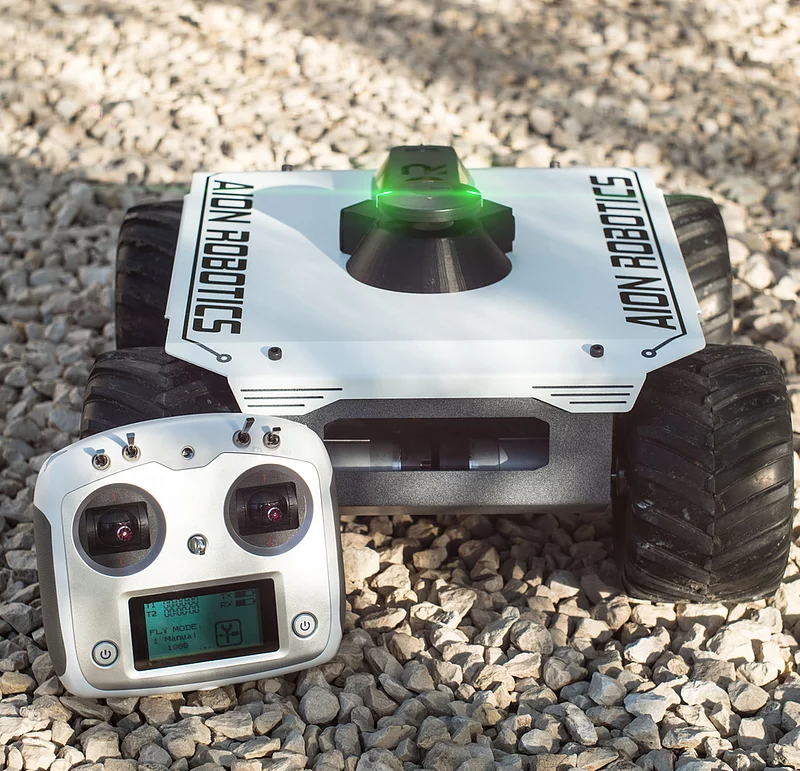}
    \includegraphics[width=0.22\columnwidth]{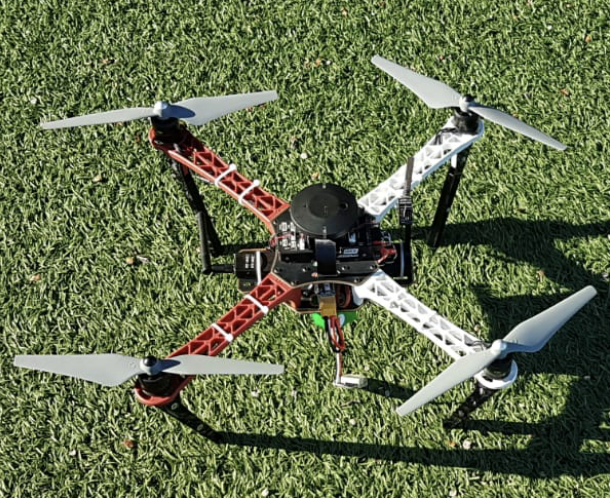}
\end{wrapfigure}
We also used two real RVs, shown to the right: 
\textbf{(Left)} Aion R1 ground rover (Aion rover)~\cite{aionr1} and 
\textbf{(Right)} Pixhawk based DIY drone (Pixhawk drone)~\cite{meier2011pixhawk}. 
Both are commodity RV systems based on the Pixhawk~\cite{pixhawk}. 
Pixhawk is a platform built on an ARM Cortex architecture that integrates ArduPilot or PX4 firmware, a flight management unit (FMU),
memory, sensors, and I/O interfaces.


To generate the firmware call graph using the \RV{}'s modified SVF, 
statically extract the required (reachable) firmware functions for each mode (\textit{analysis} phase), 
and perform instrumentation during both the \textit{analysis} and \textit{guard insertion} phases, 
we used a desktop machine equipped with an Intel Xeon E-224 CPU (4 cores, 3.4 GHz) 
running Ubuntu 22.04 (64-bit). 
After identifying the required functions for each mode and generating the instrumented firmware, 
we tested \RV{} using simulations
and  real devices.

Note that the Aion rover and the Pixhawk drone use the same platforms as ArduRover in ArduPilot 
and PXCopter in PX4, respectively; therefore, the evaluation results for ArduRover 
and the Aion rover, and for PXCopter and the Pixhawk drone, are similar.
Thus, we do not report them separately, except when evaluating the efficiency in \S\ref{subsec:efficiency}.


\noindent
\textbf{RV Missions:} To evaluate the effectiveness of \RV{}, we execute a diverse set of 40 missions (for each RV)
with varying durations and distances, randomly selected to cover a range of scenarios. 
Each mission consists of a navigation path that begins at a starting location, passes through several intermediate waypoints, and ends at a specific destination.
These missions emulate a range of real-world RV missions, including
straight-line paths for last-mile delivery drones~\cite{straighpath}, 
polygonal paths for rovers in warehouse management~\cite{polygonal}, and 
both polygonal and circular paths for agriculture and surveillance~\cite{polygoncirclular}.
The types of missions used for each RV are summarized in Table~\ref{table:missiontypes}. 

During each mission, the RV switches between different modes, 
such as \texttt{CIRCLE} and \texttt{LOITER} for ArduCopter and ArduRover, or 
\texttt{QHOVER} and \texttt{QLOITER} for ArduPlane. 
The main RVs' modes are outlined in Table~\ref{table:modeexplanation} in the Appendix.
Although our mission set includes other modes, this section presents the results for the main RV modes
that we observed are used in real-world missions (Tables~\ref{table:reduction} and \ref{table:reduction_pxcopter}).

\begin{table}[h]
    \centering
    \caption{Mission sets for each RV firmware type used to evaluate \RV{}. In the "Missions" column,
    we use SL: straight line, MW: multiple waypoints, HFE: hover at fixed elevation, CP: circular paths, PP: polygonal path.}
    \resizebox{\columnwidth}{!}{%
    \begin{footnotesize}
    \begin{tabular}{c|c|*{5}{c}}
        \multirow{2}{*}{\centering\textbf{RV (Firmware) Type}} & \multicolumn{6}{c}{\textbf{Missions}} \\
    \cline{2-7}
                     & \textbf{Total} & \textbf{SL} & \textbf{MW} & \textbf{HFE} & \textbf{PP} & \textbf{CP} \\
    \hline\hline
    \textbf{ArduCopter} &  40 & 10 &  12 & 5 & 5 & 8 \\
    \hline
    \textbf{ArduRover}  & 40 & 12 & 13 & - & 7 & 8\\
    \hline
    \textbf{ArduPlane}  & 40  & 11 & 13 & 3 & 3 & 10 \\
    \hline
    \textbf{PXCopter}  &  40 & 10 &  12 & 5 & 5 & 8 \\
\end{tabular}
    \label{table:missiontypes}
\end{footnotesize}
    }
\end{table}

\noindent
\textbf{Metrics:} 
We used four metrics to evaluate \RV{}, namely 
(1) \textit{\textbf{precision}} of the call graph, which refers to the reduction in the number of spurious call graph edges achieved 
by applying various pruning heuristics. This metric is 
similar to what prior work used~\cite{rajagopalan2023syspart,ghavamnia2020temporal}, 
(2) \textit{\textbf{missed functions}} are those that \RV{}'s dynamic profiler fails to identify 
during profiling missions using dynamic analysis. 
(3) \textit{\textbf{false positive rate (FPR)}} is the percentage of mission failures 
caused by the absence of required functions in the configuration file 
(either static or dynamic results) for the current mode, 
despite the RV needing them. 
(4) \textit{\textbf{false negative rate (FNR)}} is the percentage of functions that \RV{} 
allows during missions, even though these functions are not required by the RV's current mode. 


\subsection{Effectiveness}
\label{sec:results}
\subsubsection{\textbf{Analysis Phase}}

As discussed in \S\ref{subsub-section:static-analysis} and \S\ref{sec:implementation-analysis-phase},
\RV{} uses a modified version of SVF along with two pruning heuristics—signature-based and address-based—to generate 
and refine the firmware call graph, identifying the \textit{reachable} (\textit{required}) functions for each mode (\RVC{}).
To assess the effectiveness of these heuristics in reducing spurious edges in the call graph,
we evaluated them on ArduCopter, ArduPlane, ArduRover, and PXCopter.
As shown in Table~\ref{table:pruningreduction}, the number of edges in the original call graph generated by SVF ranges from 104K 
to 193K. By applying our pruning heuristics, 
we reduced the number of spurious edges in the call graph, achieving a precision of
approximately 36\% (66.8K),
45\% (67.1K), 48\% (60.1K), and 49\% (98.5K) for ArduCopter, ArduPlane, ArduRover and PXCopter, respectively. 
Thus, our enhancement enables \RVC{} to more effectively restrict the functions accessible in each mode, 
ensuring that it only identifies the functions actually required by each mode.
\begin{table}[h]
    \centering
    \caption{Precision after applying different heuristics. The third column shows the number of edges after applying the signature-based heuristic, and the fourth column shows the number after both heuristics.}
    \resizebox{\columnwidth}{!}{%
    \begin{tabular}{c|c|c|c|c}
    \textbf{RV (Firmware) Type}                  & \textbf{SVF}      & \textbf{+ Signature-based}  & \textbf{+ Addressed-based}  &  \textbf{Precision}        \\ 
    \hline\hline
    \textbf{ArduCopter}         &  104K & 87.4K &  66.8K   &   36\%       \\
    \hline
    \textbf{ArduRover}            & 115.3K & 80.3K  &  60.1K & 48\% \\
    \hline
    \textbf{ArduPlane}            & 122K & 86.8K & 67.1K & 45\%\\
    \hline
    \textbf{PXCopter}            & 193K & 130.2K & 98.5K & 49\%\\
    \end{tabular}
    \label{table:pruningreduction}
    }
\end{table}

The complexity of the input (i.e., firmware) affects the duration of the above process. On our platform, 
generating the call graph for each RV type (firmware) took approximately 3 to 4 hours on average, based on 10 executions of the SVF.
However, as mentioned earlier, this step is performed offline, and is done only once per firmware.


\subsubsection{\textbf{Monitoring Phase}}
\label{subsubsec:monitoring-phase}
\RV{} reduces the attack surface by restricting each mode's access to only the \textit{required} firmware functions at runtime.
In the following, we 
evaluate the effectiveness of \RV{}'s profiler in identifying the required firmware functions for each mode,  
and subsequently measure the resulting attack surface reduction achieved by \RV{}.


\noindent
\textbf{\RVA{} Profiler.}
To evaluate the \RV{} dynamic analysis output, 
we measured the number of missed functions across varying numbers of profiling and testing missions 
on different RV types, and determined how many profiling missions are needed to identify 
all the \textit{required} functions for each mode without missing any.
To do so, we used the RV missions discussed in \S\ref{subsec:experimental-setup} 
to dynamically determine the \textit{required} firmware functions for each mode and RV.
Specifically, we first randomly chose two out of the 20 test missions as profiling missions for \textit{dynamic analysis} 
for each RV. We then used the remaining 18 missions to test the identified \textit{required} functions for each mode at runtime.
We further evaluated the dynamic analysis output by gradually increasing the number of profiling missions, adding 2 missions each time to the profiling set from the testing set (e.g., 4 profiling and 16 testing missions), all the way to 20 missions. 

\begin{figure}[h]    
	\centering
	\includegraphics[scale=0.41]{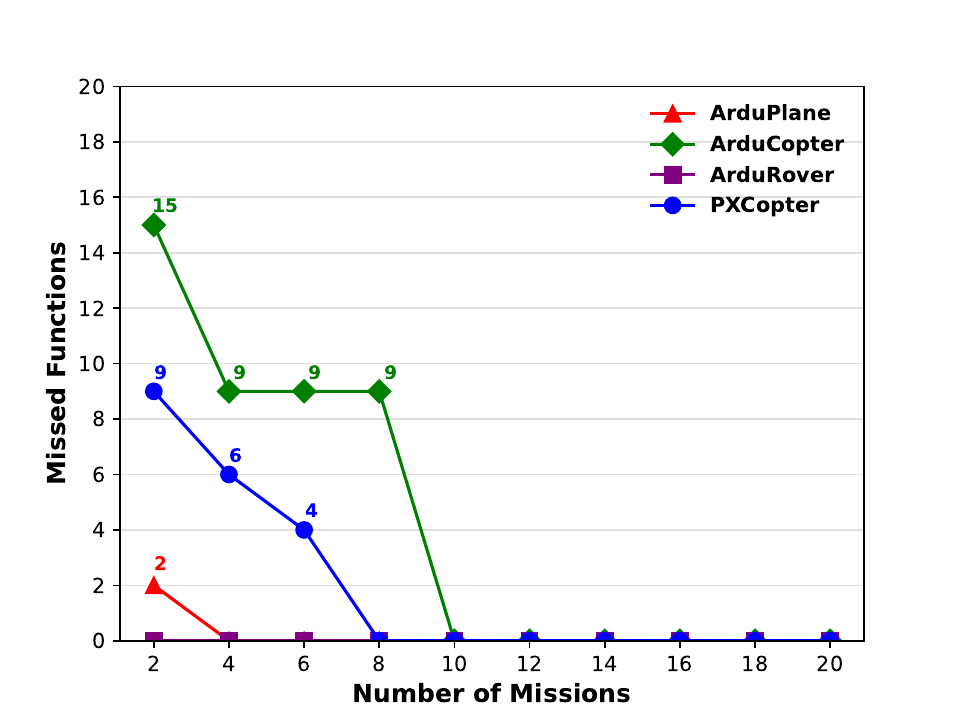}
	\caption{Number of missed functions across different missions for various platforms (i.e., ArduPlane, ArduCopter, ArduRover, and PXCopter) during dynamic analysis with \RV{}. All platforms converge to 0 missed functions after 10 missions.}
	\label{lostFunc}
\end{figure}

The results are shown in Figure~\ref{lostFunc}. 
As can be seen, all 
RVs converged to \textit{zero missed functions} after profiling within only 10 missions (i.e., 10 profiling and 10 testing missions).
Although the \RV{} profiler identified all the required functions for each mode using 10 
profiling missions, we continued the dynamic analysis evaluation by incrementally adding two new missions 
at a time to the profiling mission set 
and testing on the remaining missions (e.g., 12 profiling and 8 testing missions) 
to ensure that no missed functions appeared after convergence. 
The results confirmed that the number of missed functions for each RV remained at zero.

To further confirm this result, we evaluated the \RV{} profiler on a totally different 
 set of 20 different missions. 
The results, shown in Figure~\ref{lostFunc} (20 missions),
demonstrate that \RV{}'s dynamic profiler 
did not miss any required  functions. Therefore, all \textit{required} functions 
for each mode can be identified within 10 missions.


Although the number of profiling missions used to identify all required functions for each mode at runtime is small,
the \RV{} profiler may still miss some functions and produce false positives. 
To address this, \RVC{} can be used, 
which also includes the missing functions.
However, the device maintainer may choose to rely on the dynamic analysis results (\RVA{})
to achieve greater specialization compared to \RVC{}, which uses static analysis technique (\S\ref{sub-sec:offline-analysis} and \S\ref{sub-sec:runtime-monitor}).
Detailed results are provided below.




\noindent
\textbf{Attack Surface Reduction.} To evaluate the attack surface reduction, we used \RV{}'s dynamic and static analysis 
to identify the required firmware functions for each mode.
As we observed, after profiling 10 missions, the dynamic analysis converged to zero missing functions,  
and the number of functions remained unchanged with additional profiling.
Hence, Tables~\ref{table:reduction} and \ref{table:reduction_pxcopter} (\textit{Dynamic Analysis} row) shows the number of firmware functions required at runtime for each mode, 
as identified by \RV{}'s profiler with 10 missions.
Furthermore, the \textit{Static Analysis} row in Tables~\ref{table:reduction} and \ref{table:reduction_pxcopter} presents the number 
of \textit{reachable} functions for each mode, as determined by static analysis on the generated call graph. 
As discussed earlier (\S\ref{sub-sec:offline-analysis} and \S\ref{sub-sec:runtime-monitor}), \RVC{} and \RVA{} use the outputs of static and dynamic analysis, respectively, to specialize the firmware code.
\begin{table*}
    \centering
    \caption{Summary of firmware functions reduction (i.e., attack surface reduction through debloating) across different modes in ArduCopter, ArduPlane, and ArduRover (sub-platforms of ArduPilot). Reduction represents the percentage decrease in the number of firmware functions accessible to each mode when using \RV{} with dynamic (\RVA{}) or static analysis (\RVC{}), compared to the baseline firmware where all functions remain accessible.}
    \resizebox{\textwidth}{!}{%
    \begin{footnotesize}
    \begin{tabular}{c|c|c|c|c|c|c||c|c|c|c|c|c|c||c|c|c|c|c|c}
    \multirow{2}{*}{\diagbox{\textbf{Technique}}{\textbf{RV Type (modes)}}} 
        & \multicolumn{6}{c||}{\textbf{ArduCopter}} 
        & \multicolumn{7}{c||}{\textbf{ArduPlane}} 
        & \multicolumn{6}{c}{\textbf{ArduRover}} \\
    \cmidrule(lr){2-7} \cmidrule(lr){8-14} \cmidrule(lr){15-20}
        & \textbf{AUTO} & \textbf{CIRCLE} & \textbf{STABILIZE} & \textbf{GUIDED} & \textbf{RTL} & \textbf{LOITER} 
        & \textbf{MANUAL} & \textbf{AUTO} & \textbf{CIRCLE} & \textbf{GUIDED} & \textbf{QLOITER} & \textbf{RTL} & \textbf{QHOVER} 
        & \textbf{AUTO} & \textbf{CIRCLE} & \textbf{MANUAL} & \textbf{RTL} & \textbf{GUIDED} & \textbf{LOITER} \\
    \hline
    \textbf{Dynamic Analysis}  
        & 2869 & 3106 & 7054 & 3271 & 3153 & 2757 
        & 6891 & 2823 & 2756 & 2692 & 3120 & 2692 & 2700 
        & 2513 & 2435 & 6246 & 2433 & 2647 & 2255 \\
    \hline
    \textbf{Static Analysis}  
        & 13520 & 13450 & 13447 & 13469 & 13470 & 13452 
        & 13235 & 13253 & 13232 & 13238 & 13252 & 13239 & 13238 
        & 12637 & 12558 & 12549 & 12566 & 12583 & 12550 \\
    \hline
    \textbf{w/o \RV{}}  
        & 22653 & 22653 & 22653 & 22653 & 22653 & 22653 
        & 22514 & 22514 & 22514 & 22514 & 22514 & 22514 & 22514 
        & 20990 & 20990 & 20990 & 20990 & 20990 & 20990 \\
    \hline
    \textbf{\makecell{Reduction:    \makecell{\RVA{}\\ \RVC{}}}}
        & \makecell{87.33\% \\ 40.31\%} & \makecell{86.28\% \\ 40.62\%} & \makecell{68.86\% \\ 40.64\%} & \makecell{85.56\% \\ 40.54\%} & \makecell{86.08\% \\ 40.53\%} & \makecell{87.83\% \\ 40.61\%}
        & \makecell{69.39\% \\ 41.21\%} & \makecell{87.46\% \\ 41.13\%} & \makecell{88.75\% \\ 41.22\%} & \makecell{88.04\% \\ 41.20\%} & \makecell{86.14\% \\ 41.14\%} & \makecell{88.04\% \\ 41.19\%} & \makecell{88\% \\ 41.20\%}
        & \makecell{88.02\% \\ 39.80\%} & \makecell{88.40\% \\ 40.17\%} & \makecell{70.24\% \\ 40.21\%} & \makecell{88.40\% \\ 40.13\%} & \makecell{87.39\% \\ 40.05\%} & \makecell{89.26\% \\ 40.20\%} \\
    \end{tabular}
    \label{table:reduction}
    \end{footnotesize}
    }
\end{table*}

\begin{table}[h]
    \centering
    \caption{Summary of firmware functions reduction in PXCopter across different modes (continuation of Table~\ref{table:reduction}).}
    \resizebox{\columnwidth}{!}{%
    \begin{footnotesize}
    \begin{tabular}{c|c|c|c|c}
    \textbf{Technique} & \textbf{TAKEOFF} & \textbf{MISSION} & \textbf{LOITER} & \textbf{RTL} \\
    \hline
    \textbf{Dynamic Analysis}  & 4320 & 2765 & 2660 & 2853 \\
    \hline
    \textbf{Static Analysis}   & 12244 & 12238 & 12208 & 12227 \\
    \hline
    \textbf{w/o \RV{}}         & 20805 & 20805 & 20805 & 20805 \\
    \hline
    \textbf{\makecell{Reduction:    \makecell{\RVA{}\\ \RVC{}}}}
        & \makecell{79.23\% \\ 41.15\%} & \makecell{86.71\% \\ 41.18\%} & \makecell{87.21\% \\ 41.32\%} & \makecell{86.29\% \\ 41.23\%} \\
    \end{tabular}
    \label{table:reduction_pxcopter}
    \end{footnotesize}
    }
\end{table}


The results reveal that, on average, 85\% and 41\% of firmware functions are not required according to dynamic and static analysis, respectively. 
In particular, when \RV{} uses static analysis outputs, 
it reduces the firmware functions by an average of 41\% (Reduction \RVC{} in Tables~\ref{table:reduction} and \ref{table:reduction_pxcopter}), 
while achieving a 0\% \textbf{FPR} 
and an average \textbf{FNR} of 44\%, as this technique tends to overestimate the functions required for each mode (\S\ref{sub-sec:offline-analysis}).
Attempts to access debloated functions 
lead to activating \texttt{Fail-Safe} mode,  and raise an alarm
to the device maintainer 
(\S\ref{sub-sec:runtime-monitor}).

On the other hand, when the device maintainer uses \RVA{}, 
which uses dynamic analysis—generated by profiling 
10 missions—no missions fail, resulting in a 0\% \textbf{FPR} and an 85\% reduction 
in the attack surface (Reduction \RVA{} in Tables~\ref{table:reduction} and \ref{table:reduction_pxcopter}). 
Consequently, any attempt to access the debloated functions triggers a switch of the RV  
to \texttt{Fail-Safe} mode, thereby restricting the attacker's access to only 15\% of the firmware's functions with \RVA{}. 
Furthermore, all functions allowed by \RVA{} are actually required during execution 
(identified by profiling different missions), resulting in a 0\% \textbf{FNR}.

From the results (Table~\ref{table:reduction} and \ref{table:reduction_pxcopter}), we can infer two main trends. 
First, each mode requires only a subset of the firmware functions rather than the entire firmware. 
For instance, when the RV is in \texttt{AUTO} mode, it requires approximately 2800 
firmware functions, whereas in \texttt{RTL} mode, 
it requires a different set of around 3000 firmware functions (addressing L1, L2, and L4 in \S\ref{sub-sec:current-limitations}).
Second, because RVs switch between different modes, including returning to earlier modes,
they need an adaptive debloating technique that is reversible, in contrast to previous
debloating techniques~\cite{ghavamnia2020temporal,rajagopalan2023syspart} that use irreversible mechanisms to disable unwanted features (addressing L5 in \S\ref{sub-sec:current-limitations}).

\noindent
\textbf{Attack Evaluation.} According to the list of most dangerous software weaknesses from CWE~\cite{cweweb}, out-of-bounds write is one 
of the most prevalent vulnerabilities. 
Thus, to evaluate \RV{}'s effectiveness (\RVA{}, \RVC{}) in restricting firmware access,
we injected a buffer-overflow vulnerability—inspired by real-world vulnerabilities~\cite{stackoverflow,bufferoverflow}—into 
a function that processes user input, following prior work~\cite{li2025software,kim2018securing}.
An attacker who exploits this vulnerability can 
hijack control flow and perform a code-reuse attack.
In our evaluation, we exploited this vulnerability to hijack the control flow and then 
executed three sensitive firmware functionalities, such as \texttt{disarm\_motors} (\S\ref{sub-sec:current-limitations}), while the device was not in the 
correct mode. We discuss the three attacks in detail. 

\textbf{\textit{(A1) Disarming RVs.}} ArduCopter provides 
a safety-critical function, \texttt{disarm\_motors} (Listing~\ref{list}), which disarms the 
motors during benign tasks such as flipping the device upright in \texttt{Turtle} mode (\S\ref{sub-sec:current-limitations}). 
However, after hijacking the control flow,
our attack reused the \texttt{disarm\_motors} function during a mission
(e.g., while the device is in \texttt{GUIDED} mode), causing the device to crash.
The attack succeeded because the firmware allowed unrestricted access to any function, 
including \texttt{disarm\_motors}, regardless of the current mode.
In contrast, \RV{} detected the attack and switched the device mode to \texttt{Fail-Safe}, 
allowing the device to land safely or return to the launch point (\S\ref{sub-sec:overview}), 
because \texttt{disarm\_motors} is restricted and not 
accessible from other modes, such as \texttt{GUIDED}. 
The \textit{disarming} functionality also exists in other platforms, such as ArduPlane and ArduRover, 
and is used during specific events like parachute release or crash detection.
An attacker can perform the same attack to disarm the motors and cause physical damage.
Again, \RV{} detected such attacks because disarming should not be allowed in modes such as \texttt{QHOVER}.

\textbf{\textit{(A2) Stopping Motors.}} ArduCopter and ArduPlane provide a function called \texttt{output\_min},
which stops all motors and cuts engine power when specific events occur, such as when the device has landed. 
Similar to the previous attack, our stopping motors attack hijacks the control flow of the victim device (firmware)
and calls \texttt{output\_min} to stop the motors at an incorrect time (e.g., during mission while the device is in \texttt{GUIDED}),
causing the device to crash. In our experiments, we confirmed that an attacker can easily disrupt a mission 
and damage the device by invoking this function. 
This attack was also detected and prevented by \RV{},
since \texttt{output\_min} is not reachable from modes such as \texttt{GUIDED} or \texttt{QHOVER}.

\begin{figure}[h]
    \centering
    \begin{subfigure}[b]{0.38\columnwidth}
        \centering
        \includegraphics[width=\columnwidth]{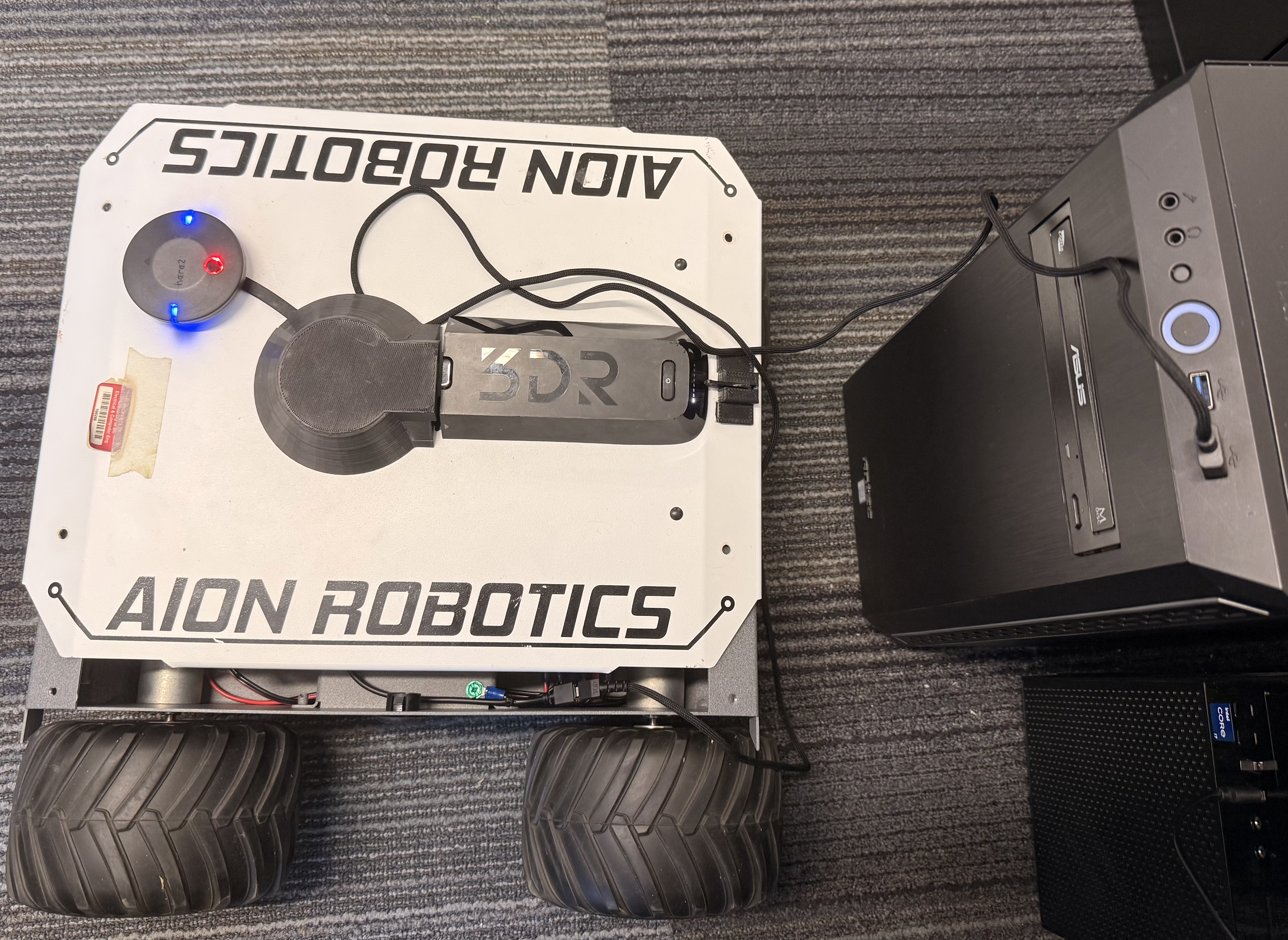}
        \caption{Boot Firmware}
        \label{fig:realsetup1}
    \end{subfigure}
    \hspace{0.01\columnwidth}
    \begin{subfigure}[b]{0.38\columnwidth}
        \centering
        \includegraphics[width=0.9\columnwidth]{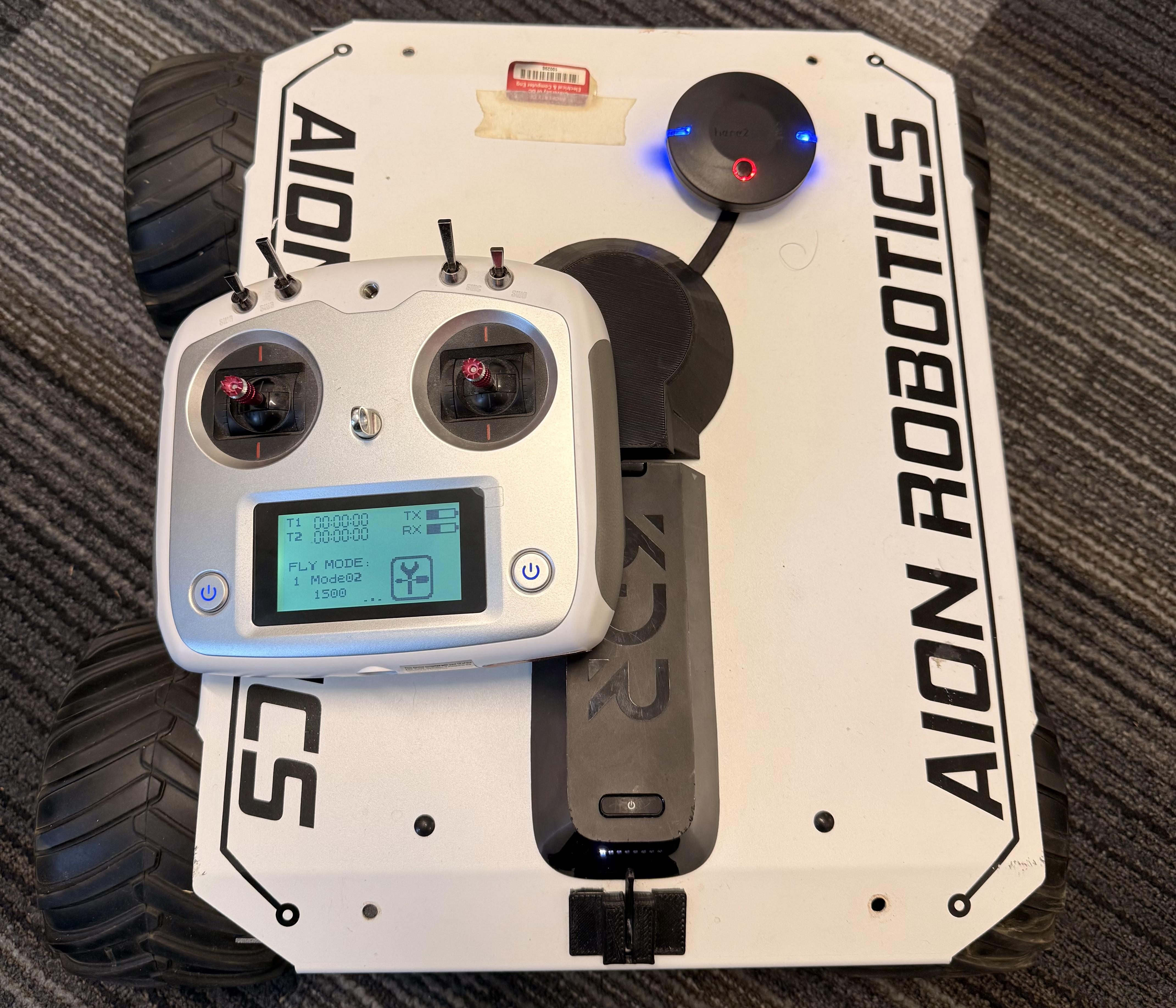}
        \caption{Runtime}
        \label{fig:realsetup2}
    \end{subfigure}
    \caption{Real-world attack (A3) on Aion Rover.}
    \label{fig:realsetup}
\end{figure}

\textbf{\textit{(A3) Disarming Aion Rover.}} This attack case demonstrates the effectiveness of \RV{} 
on a real-world device, the Aion Rover, discussed in \S\ref{subsec:experimental-setup}.
Similar to the previous attack cases, a buffer overflow vulnerability is intentionally inserted into the 
function that processes user input, inspired by previous work~\cite{li2025software}. During the mission, this vulnerability is triggered, 
allowing the attacker to hijack the control flow and invoke a safety-critical function, \texttt{disarm}, 
from an incorrect mode (i.e., \texttt{MANUAL}).
As a result, the motors are disarmed, and the rover stops abruptly, which may cause collisions with other nearby rovers.
However, \RV{} successfully detected this attack because the \texttt{disarm} function is not included in the list of required functions for modes such as \texttt{MANUAL}.
Figure~\ref{fig:realsetup} shows the setup for this attack on the Aion Rover in our lab.

\subsection{Efficiency}
\label{subsec:efficiency}
\subsubsection{\textbf{Performance and Power Overhead}} 
As discussed in \S\ref{sub-sec:overview}, the only component of \RV{} that incurs overhead on the RV is the monitoring component, 
which monitors mode-switching function and checks indirect control flow transfer instructions to specialize the firmware code for each mode.
Therefore, since both \RVC{} and \RVA{} use this component, their results are similar.
To evaluate the performance overhead, we used the real RVs mentioned in \S\ref{sec:implementation}, 
i.e., the Aion rover and Pixhawk drone, as performance overheads in simulated RVs are dependent on the computing platform used for simulation. 
We used the scheduler in the RV autopilot platforms (ArduPilot and PX4) 
to track total CPU time and measured \RV{}'s overhead by analyzing the additional CPU time 
incurred when \RV{} was deployed on the RV.

We found that \RV{} incurs modest  performance overheads of 3.5\% on the Aion rover 
and 4.3\% on the Pixhawk drone, with an average of 3.9\%.
We also found that there was \textit{no increase} in the overall mission time of the RVs with \RV{} for the missions discussed in \S\ref{subsec:experimental-setup}.

Furthermore, we estimated the power consumption of \RV{} running on those RVs, 
which are powered by a 5000 mAH battery.
Since the processor typically accounts for 12\% of the total power consumption~\cite{powerconsuption} 
and \RV{} adds an average overhead of 3.9\%, we estimate that the RV's total power and energy consumption increases by approximately 0.47\%, which is negligible. 

\subsubsection{\textbf{Memory Overhead}}

To measure \RV{}'s memory overhead, only the configuration files generated during the \textit{analysis} phase
and the instrumented firmware produced during the \textit{guard insertion} phase are stored on the device. 
Thus, we focus on the overhead introduced by 
these two phases, as they are the only ones that incur overhead to memory usage on the device.
As discussed in \S\ref{subsec:implementation-guard} and \S\ref{subsec:monitoring-phase},
\RV{} builds the firmware using the LLVM compiler toolchain (i.e., gclang, gclang++, and clang). 
This process involves extracting the bitcode from the original executable firmware file 
and then generating a new executable file from the instrumented bitcode for deployment on the device. 

\begin{table}[H]
    \centering
    \caption{Firmware size with and without using \RV{} guard insertion phase.}
    \resizebox{\columnwidth}{!}{%
    \begin{tabular}{c|c||c||c||c}
    \diagbox{\textbf{Technique}}{\textbf{RV Type}}  & \textbf{ArduCopter} & \textbf{ArduPlane} &\textbf{ArduRover}&\textbf{PXCopter}\\ 
    \hline
    \textbf{w/o \RV{}}   & 6.2 MB  & 6.1 MB  &  5.7 MB & 6.2 MB\\
    \textbf{\RV{}}       & 6.5 MB & 6.4 MB &  5.9 MB & 6.4 MB\\
    \hline
    \textbf{Overhead}       & \textbf{0.3 MB (4.8\%)} & \textbf{0.3 MB (4.9\%)} & \textbf{0.2 MB (3.5\%)} & \textbf{0.2 MB (3.2\%)}\\
    
    \end{tabular}
    \label{table:memoverhead}
    }
\end{table}

We find that the instrumented firmware introduces an average memory overhead of 0.25 MB 
across different platform types, 
corresponding to approximately a 4\% increase over the original firmware.
A summary of firmware sizes is presented in Table~\ref{table:memoverhead}. 
Further, the mode-based configuration files generated by \RVC{} and \RVA{} (\S\ref{sub-sec:offline-analysis}) incur 
an average memory overhead of 0.65 MB and 0.17 MB, respectively.
These files are not included in the firmware and thus do not contribute to the 4\% size increase.
Overall, the total memory overhead (i.e., instrumented firmware+configuration files) is negligible ($<$ 4\%), as the memory available on the real RVs is over 20 MB.

Note that, like other debloating (specialization) techniques~\cite{abubakar2021shard,ghavamnia2020temporal,kim2018securing} that retain the firmware code while restricting access,
\RV{} also does not reduce the memory usage. 
As discussed earlier, mode-based embedded devices require the entire firmware, 
but only the functions relevant to the current mode should be accessible in it.

    

\label{evaluatesec}

\section{Related Work}

\textbf{Fixed debloating techniques} fall into two types based on whether they \textit{retain} 
or \textit{remove} unneeded features, such as security-critical system calls or hardware features.

\textit{1. Retaining Features.}
Abubakar et al.~\cite{abubakar2021shard} proposed SHARD, which leverages context-aware hardening to specialize 
the Linux kernel at the application and system call levels. 
SHARD uses both static and dynamic analysis to identify the required kernel code 
for executing the system call invoked by an application. 
Afterwards, at runtime, SHARD ensures that only the identified kernel code is allowed to execute 
when the application calls the same system call.
However, SHARD switches all kernel code pages for each system call invocation, 
making it unsuitable for resource-constrained embedded devices like RVs.


Minion~\cite{kim2018securing} proposed a thread-level memory isolation technique for real-time microcontroller systems (MCS) 
that isolates the memory space for each process through offline static analysis of firmware 
and runtime memory access control using the Memory Protection Unit (MPU). 
However, Minion remains vulnerable to control flow hijacking attacks, 
as attackers can exploit memory corruption vulnerabilities in the shared memory between processes 
to hijack control flow and construct ROP or JOP chains.
Further, since Minion uses the MPU to enforce memory isolation, 
it applies clustering techniques to group functions. 
However, this may allow processes to access functions that they do not require (\S\ref{sub-sec:current-limitations}).


Gu et al. proposed FACE-CHANGE~\cite{gu2014face}, a kernel debloating technique that employs dynamic analysis 
to generate kernel profiles for each application, identifying the required kernel code at the application level. 
However, FACE-CHANGE's reliance on runtime profiling leads to incomplete identification of required kernel code 
(i.e., customized kernel view), limiting its analysis to the subset of the program executed during profiling.
Furthermore, FACE-CHANGE suffers from coarse specialization granularity (i.e., single kernel view for the entire application),
and hence attackers can still hijack the control flow of the sensitive programs (e.g., flight control program in RVs)
and manipulate the devices' behavior.


\textit{2. Removing Features.}
Hu et al. proposed IRQDebloat~\cite{hu2022irqdebloat}, a debloating technique that leverages dynamic analysis to identify interrupt handlers 
and disable undesired hardware functionalities by rewriting their handlers with a no-op function in the target firmware. 
Hu et al. proposed Hacksaw~\cite{hu2023hacksaw}, a kernel specialization technique 
that uses a list of hardware components attached to the target machine, called the device inventory, 
to remove kernel device drivers related to unnecessary hardware components.

Wu et al. proposed LightBlue~\cite{wu2021lightblue}, a profile-aware debloating technique that uses static analysis 
to identify the Bluetooth profiles (e.g., Advanced Audio Distribution Profile (A2DP)) 
utilized by an application (e.g., an Android app). Then, it removes unused code from the Bluetooth stack, 
including Bluetooth host code and firmware, thereby reducing the attack surface. 
However, LightBlue is limited to debloating of Bluetooth stacks and does not consider other critical features, 
such as firmware functionalities (e.g., \texttt{disarm\_motors} in RV autopilot software).


Jahanshahi et al. proposed Minimalist~\cite{jahanshahi2023minimalist}, a semi-automated debloating technique
for PHP web applications (e.g., WordPress and PhpMyAdmin) that uses static analysis 
to generate a call graph for these applications. 
Minimalist then removes unreachable and unused code 
based on the generated call graph and recorded access-log files, 
which include users' interactions with the web applications.
However, Minimalist has two limitations as follows. 
First, users need to interact with the web server for each application to generate the access-log files, 
which is both time-consuming and potentially incomplete.
Second, since Minimalist relies on user requests, it may remove essential parts of the code 
that are not captured in the log files, potentially causing system crashes.

All of these debloating techniques, which \textit{remove unwanted features} such as hardware functionalities 
or kernel code, are irreversible, meaning that once unneeded features are removed, 
they cannot be restored.
However, embedded devices may require different functionalities during various phases of execution.
Therefore, these techniques are only suitable 
when certain features are guaranteed to remain unused during runtime; hence, our work is orthogonal 
to this line of work, complementing these techniques. 

\textbf{Adaptive Debloating Techniques}
As mentioned in \S\ref{sec-intro}, \textit{fixed debloating techniques} do not consider changes in application requirements
at runtime, and as a result, they cannot disable critical features, such as the \texttt{fork} and \texttt{execve} system calls,
which are necessary during the initial stages of an application.
%
To address this issue, Ghavamnia et al.~\cite{ghavamnia2020temporal} and Rajagopalan et al.~\cite{rajagopalan2023syspart}
proposed adaptive system call specialization techniques for server applications, 
which are discussed in detail,in Sections \S\ref{sec-intro} and \S\ref{sub-sec:current-limitations}.

Unlike existing adaptive debloating techniques that rely on irreversible mechanisms 
to disable system calls, \RV{} dynamically specializes firmware code
at runtime for each mode. This is particularly important for mode-based embedded devices, 
such as RVs, where firmware functions may need to be re-enabled depending 
on the active mode (\S\ref{subsubsec:monitoring-phase}). Further, while prior techniques focus solely on system 
call disabling, \RV{} provides function-level specialization of firmware.

\label{releatedsec}

\section{Conclusion and Future Work}

We propose \RV{},
the first \DT{} debloating framework for \textit{mode-based} embedded devices.
\RV{} addresses limitations of existing debloating techniques by using static or dynamic analysis
to \textit{automatically} identify the \textit{required} firmware functions for each mode and 
reduce the attack surface by restricting access to unneeded functions at runtime.
Furthermore, \RV{} uses a software-only  technique to track mode-switches and specialize the firmware code at runtime.
\RV{} is implemented using the LLVM compiler, making it portable across  platforms.

We evaluated \RV{} using real missions with various scenarios on six different RVs running 
either ArduPilot or PX4,
two of the most widely used RV firmware.
The results reveal that 
\RV{} can reduce the RVs' attack surface by an average of 41\% with static analysis and 85\% with dynamic analysis.
Furthermore, \RV{} incurred a performance overhead of 3.9\%, and negligible power and memory overheads on those RVs ($<$ 4\%). 
Finally, \RV{} successfully mitigated three attacks on the RVs.

We outline one possible direction to be explored in future studies.
Although we evaluated \RV{} using either static or dynamic analysis, 
these techniques can be combined as a hybrid approach. 
\textit{\RV-hybrid} can first check using the dynamic output; if the target is not found, 
it then checks the static output. Since static analysis tends to overestimate, 
control flow integrity (CFI) techniques can then be applied using the generated call graph  
to ensure that all control flow transfers adhere to the CFG.

\label{conclusionsec}




\bibliographystyle{IEEEtran}
\bibliography{main}

\appendices

\section{Implementation Details}
\label{seca:implementation-details}

A summary of the \RV{} implementation across its five main components, 
corresponding to the different phases (\S\ref{sec:implementation}, \S\ref{sec:implementation-analysis-phase}, 
\S\ref{subsec:implementation-guard}, and \S\ref{subsec:monitoring-phase})
is presented in Table~\ref{table:LOCcomponent}.

\begin{table}[H]
    \centering
    \caption{Lines of code in \RV{} components.}
    \begin{tabular}{c|c}
    \textbf{Component}                  & \textbf{Lines of Code}               \\ 
    \hline\hline
    \textbf{Pruning Heuristics}         & 220 C++                     \\
    \textbf{Instrumentation}            & 1000 C++ (two LLVM passes)  \\
    \textbf{\RV{} Profiler}                   & 500 C                       \\ 
    \hline
    \textbf{Guard Insertion} & 1100 C++ (two LLVM passes)  \\
    \hline
    \textbf{\RV{} Monitor}                    & 400 C                       \\ 
    \hline
    \textbf{Total}                     & 3220                       
    \end{tabular}
    \label{table:LOCcomponent}
\end{table}

\subsection{Analysis Phase}
\label{sec:implementation-analysis-phase}
As mentioned earlier, the instrumentation component of the analysis phase takes the firmware source code as input.
In the first step, the firmware is compiled to LLVM bitcode (\textit{bc} file) using the \textit{gclang} and \textit{gclang++} compilers~\cite{gllvm}.
Then, if \RVC{} is selected, it uses the modified version of SVF with two pruning heuristics—signature-based and address-based—to generate a pruned call graph. 
We modified the TSCP implementation~\cite{ghavamnia2020temporal} by adding new heuristics to refine the call graph.
A Python script then analyzes this call graph to generate lists of reachable functions for each mode through static analysis.

If \RVA{} is selected, the instrumentation component links the firmware bitcode with \RV{}'s profiler functions (i.e., \textit{mode\_entry} and \textit{log\_fn}). 
This component then takes the mode-switching function as input and iterates through the firmware's instructions.
Specifically, it inserts \textit{mode\_entry} at locations within the mode-switching function where the return value is \texttt{true}, indicating a successful mode change. 
Also, it inserts \textit{log\_fn} at the entry points of other firmware functions.
\begin{lstlisting}[caption= Inserting \textit{log\_fn} at the entry of functions. ,label=listanalysis ,language=myLangLL]
define dso_local zeroext i1 @AP_Arming_Plane(%class.AP_Arming_Plane*, i32, i1 zeroext) #0 align 2 !dbg !97833 {
// Allocation Variables
call void @log_fn(arguments)
// ... omit
}
\end{lstlisting}

In the example shown in Listing~\ref{listanalysis}, the instrumentation component inserts \textit{log\_fn}
at the beginning of the function, after the allocation instructions.
Finally, \RV{} automatically executes the instrumented firmware with different missions 
in a benign environment and tracks its execution using profiler functions to identify the required functions 
for each mode at runtime.

\subsection{Guard Insertion Phase}
\label{subsec:implementation-guard}

In the next phase, a different instrumentation component takes the firmware and mode-switching function(s) as input.
Once the firmware is compiled to LLVM bitcode using \textit{gclang} and \textit{gclang++} compilers and linked with \RV{}'s monitor functions 
(i.e., \textit{mode\_entry\_runtime} and \textit{monitor\_fn}), \RV{} first identifies 
the mode-switching function(s). It then inserts \textit{mode\_entry\_runtime}, which monitors mode changes and loads 
the required functions for the new mode from the configuration file (either static or dynamic output) at
locations where the mode successfully changes. Part of the \textit{mode\_entry\_runtime} function in the LLVM IR format is shown in Listing~\ref{llvmex}. Next, the component iterates over the firmware functions and replaces indirect control flow transfer instructions (e.g., indirect call sites represented by \textit{CallInst} in LLVM IR) and
return instructions (\textit{ReturnInst} in LLVM IR) with \textit{monitor\_fn}, which controls target execution, 
raises an alarm, and switches to \texttt{Fail-Safe} mode if needed.

\begin{lstlisting}[caption= Inserting \textit{mode\_entry\_runtime} before return instruction (Listing~\ref{instrumentation_example}).,label=listmonitor ,language=myLangLL]
1 define dso_local zeroext i1 @Plane18set_mode_by_number(%class.Plane*, i8 zeroext, i8 zeroext) #0 align 2 !dbg !566193 {
// ... omit
2 call void @mode_entry_runtime(arguments)
3 ret i1 %22, !dbg !566216
4 }
\end{lstlisting}

For instance, this component inserts \textit{mode\_entry\_runtime} before the return instruction 
when its value is true, as shown in Listing~\ref{listmonitor}.
Finally, it stores the instrumented file.


\subsection{Monitoring Phase}
\label{subsec:monitoring-phase}

In the final phase, \RV{} builds the instrumented firmware executable using the \textit{Clang} compiler~\cite{clangurl}.
At runtime (during an RV mission), the \RV{} monitor is triggered when a mode change occurs. 
The monitor component then loads the corresponding configuration file, which includes the required functions for the new mode identified by either static or dynamic analysis.
Furthermore, the \RV{} monitor is triggered whenever an indirect control flow transfer or a return instruction is executed. 
In such cases, the component checks permissions based on the loaded configuration for the current mode and logs the event, 
raising an alarm or switching to \texttt{Fail-Safe} mode if an unauthorized execution is detected. 

\section{LLVM IR Example}
\label{sec-app:llvmir}

\begin{figure}[h]
    \lstset{basicstyle=\scriptsize}
\begin{lstlisting}[caption=A part of the LLVM IR for the \textit{mode\_entry\_runtime} function in \RV{}'s mode-switching monitor (line 2 of Listing~\ref{listmonitor}).,label=llvmex]
1 define dso_local void @mode_entry_runtime(i8 zeroext) #0 {
2    %2 = alloca i8, align 1
3    %3 = alloca %struct.FuncEntry*, align 8
4    %4 = alloca %struct.FuncEntry*, align 8
5    %5 = alloca %struct.UT_hash_handle*, align 8

6  // other alloca instructions (omit)

7    store i8 %0, i8* %2, align 1
8    %44 = load i32, i32* @curr_mode_id, align 4
9    %45 = load i8, i8* %2, align 1
10   %46 = zext i8 %45 to i32
11   %47 = icmp eq i32 %44, %46
12   br i1 %47, label %48, label %49
  
13  ; <label>:48:                              ; preds = %1
14   br label %2248
  
15  ; <label>:49:                              ; preds = %1
16   store volatile i32 1, i32* @mode_switching, align 4
17   %50 = load %struct.FuncEntry*, %struct.FuncEntry** @allowed_functions, align 8
18   %51 = icmp ne %struct.FuncEntry* %50, null
19   br i1 %51, label %52, label %437
  
20  ; <label>:52:                              ; preds = %49
21   %53 = load %struct.FuncEntry*, %struct.FuncEntry** @allowed_functions, align 8
22   store %struct.FuncEntry* %53, %struct.FuncEntry** %3, align 8
23   %54 = load %struct.FuncEntry*, %struct.FuncEntry** @allowed_functions, align 8
24   %55 = icmp ne %struct.FuncEntry* %54, null
25   br i1 %55, label %56, label %61
  
26  // other instructions (omit)

27  ; <label>:2248:              ; preds = %2245, %495, %453, %48
28   ret void
29 }
\end{lstlisting}
\end{figure}

Listing~\ref{llvmex} shows a part of the \textit{mode\_entry\_runtime} function in LLVM IR format, 
which is used by \RV{} to instrument mode-switching function(s) in \textit{guard insertion} phase.
For the sake of simplicity, we only present a partial snippet of this function. 
As previously discussed in \S~\ref{sub-sec:offline-instrumentation-monitoring} and \S\ref{subsec:implementation-guard}, 
before executing the device and monitoring it for specialization, 
\RV{} instruments the mode-switching function(s) during the \textit{guard insertion} phase.
\RV{} iterates over these functions to identify locations where 
a mode switch is successful (e.g., returning true or a new mode), and then
inserts a call to \textit{mode\_entry\_runtime} at those points. 
The purpose of \textit{mode\_entry\_runtime} function
is to monitor mode changes at runtime and load the configuration file corresponding to the new mode, allowing
the \RV{} monitor to specialize and restrict access to other firmware function via another function called 
\textit{monitor\_fn}, which controls indirect control flow transfer instructions. 

As can be seen in the Listing~\ref{llvmex}, after initializing the variables (lines 2 to 7), 
the function compares the 
current mode with the new mode passed as an argument (e.g., \textit{set\_mode\_by\_number} function in Listing~\ref{listmonitor})
in lines between 7 to 12, to ensure they are not the same.
Then, in lines 16 to 25, it loads the new list of functions and their corresponding addresses into a structure (e.g., \texttt{allowed\_functions}). 
Once this process is complete, control returns to the mode-switching function (line 28).

\section{ArduPilot \& PX4 Modes}
\label{sec-app:modesnames}

The explanation of the main modes of ArduPilot and PX4 across different RV types (firmware), including ArduCopter, ArduPlane, ArduRover, and PXCopter, 
is summarized in Table~\ref{table:modeexplanation}. For a more detailed explanation and information on other modes, 
please refer to the ArduPilot and PX4 documentations~\cite{ardupilotdoc,px4doc}.

\begin{table*}[]
    \centering
	\caption{ArduPilot and PX4 Mode Details. For details about each mode, see the ArduPilot and PX4 documentations~\cite{ardupilotdoc,px4doc}.}
    \resizebox{\textwidth}{!}{%
    \begin{footnotesize}
    \begin{tabular}{|c|c|c|}
    \hline
    \textbf{RV Type} & \textbf{Mode} & \textbf{Mode Description} \\ \hline
    \multirow[c]{6}{*}{\textbf{ArduCopter}}  & \textbf{AUTO} & Automatically flying to a pre-defined mission \\
             & \textbf{CIRCLE} & Automatically circling a point in front of the RV \\
             & \textbf{STABILIZE} &  Self levels in the roll and pitch axis\\
             & \textbf{GUIDED} & Flying to points commanded by GCS \\
             & \textbf{RTL} & Returning to home (launch) location, can also include landing \\
             & \textbf{LOITER} & Holding altitude and position, uses GPS for movements \\
             & \textbf{TURTLE} & Flipping an inverted vehicle upright \\
    \hline\hline
    \multirow[c]{7}{*}{\textbf{ArduPlane}} & \textbf{MANUAL} &  Regular RC control, no stabilization\\
              & \textbf{AUTO} & Flying to a preloaded mission (set of GPS coordinates)\\
              & \textbf{CIRCLE} &  Gently turning the RV around the point  \\
              & \textbf{GUIDED} &  Flying and circling the RV to points commanded by GCS\\
              & \textbf{QLOITER} &  Automatically attempting to maintain the current location, heading and altitude\\
              & \textbf{RTL} & Returning to home (launch) location \\
              & \textbf{QHOVER} &  Maintaining a consistent altitude while allowing roll, pitch, and yaw to be controlled normally\\
    \hline\hline
    \multirow[c]{6}{*}{\textbf{ArduRover}} & \textbf{AUTO} & Following a preloaded mission (set of GPS coordinates)\\
               & \textbf{CIRCLE} & Automatically orbits a point, sent from the GCS, located in front of the RV\\
               & \textbf{MANUAL} & Manually control the vehicle's throttle and steering \\
               & \textbf{RTL} &  Returning to home (launch) location \\
               & \textbf{GUIDED} & Controlling and moving the RV to points commanded by GCS\\
               & \textbf{LOITER} & Maintaining the current location and heading\\
    \hline\hline
    \multirow[c]{4}{*}{\textbf{PXCopter}} & \textbf{TAKEOFF} & Taking off vertically and then switches to Hold mode\\
               & \textbf{MISSION} & Following a preloaded mission (set of GPS coordinates)\\
               & \textbf{LOITER} & Holding altitude and position, uses GPS for movements \\
               & \textbf{RTL} &  Returning to home (launch) location and landing \\
    \hline
    \end{tabular}
    \end{footnotesize}
    }
    \label{table:modeexplanation}
\end{table*}

\end{document}